\documentclass[a4paper,fleqn,usenatbib,useAMS]{mnras}

\usepackage{newtxtext,newtxmath}
\usepackage[T1]{fontenc}
\usepackage{ae,aecompl}

\usepackage{graphicx}	% Including figure files
\usepackage{amsmath}	% Advanced maths commands
\usepackage{amssymb}	% Extra maths symbols
\usepackage{braket}
\usepackage{multirow}
\usepackage{threeparttable}

\newcommand{\vv}{\upsilon}

\newcommand{\Lum}{\mathcal{L}}
\newcommand{\eps}{\varepsilon}
\newcommand{\CR}{\rmn{cr}}

\usepackage[normalem]{ulem}
\usepackage[usenames]{xcolor}

\def\del#1{{}}

\def\C#1{#1}

\defcitealias{Jacob2016a}{JP17}

\title[Cosmic ray heating in cool core clusters II]{Cosmic ray heating in cool
  core clusters -- II. Self-regulation cycle and non-thermal emission}

\author[S.~Jacob and C.~Pfrommer]{
Svenja Jacob$^{1,2}$\thanks{svenja.jacob@h-its.org (SJ), christoph.pfrommer@h-its.org (CP)} and
Christoph Pfrommer$^{1}$\footnotemark[1]
\\
$^{1}$Heidelberg Institute for Theoretical Studies, Schloss-Wolfsbrunnenweg 35, D-69118 Heidelberg, Germany \\
$^{2}$Zentrum f\"ur Astronomie der Universit\"at Heidelberg, Astronomisches Recheninstitut, M\"onchhofsstr. 12-14, D-69120 Heidelberg, Germany\\
}

\date{Accepted XXX. Received YYY; in original form ZZZ}

\pubyear{2016}

\begin{document}
\label{firstpage}
\pagerange{\pageref{firstpage}--\pageref{lastpage}}
\maketitle

% Abstract of the paper 250 words
\begin{abstract}
  \noindent
  Self-regulated feedback by active galactic nuclei (AGNs) appears to be
  critical in balancing radiative cooling of the low-entropy gas at the centres
  of galaxy clusters and in regulating star formation in central galaxies. In a
  companion paper, we \C{found steady-state solutions} of the hydrodynamic
  equations that are coupled to the cosmic ray (CR) energy equation for a large cluster
  sample. In those solutions, radiative cooling in the central region is
  balanced by streaming CRs through the generation and dissipation of resonantly
  generated Alfv{\'e}n waves and by thermal conduction at large radii. Here, we
  demonstrate that the predicted non-thermal emission resulting from hadronic CR
  interactions in the intra cluster medium exceeds observational radio (and
  gamma-ray) data in a subsample of clusters that host radio mini halos
  (RMHs). In contrast, the predicted non-thermal emission is well below
  observational data in cooling galaxy clusters without RMHs. These are
  characterized by exceptionally large AGN radio fluxes, indicating high CR
  yields and associated CR heating rates. We suggest a self-regulation cycle of
  AGN feedback in which non-RMH clusters are heated by streaming CRs
  homogeneously throughout the central cooling region.  We predict {\em radio
    micro halos} surrounding the AGNs of these CR-heated clusters in which the
  primary emission may predominate the hadronically generated emission. Once the
  CR population has streamed sufficiently far and lost enough energy, the
  cooling rate increases, which explains the increased star formation rates in
  clusters hosting RMHs. Those could be powered hadronically by CRs that have
  previously heated the cluster core.
\end{abstract}


%%%%%%%%%%%%%%%%%%%%%%%%%%%%%%%%%%%%%%%%%%%%%%%%%%

%%%%%%%%%%%%%%%%% BODY OF PAPER %%%%%%%%%%%%%%%%%%

\section{Introduction}

The central cooling time of the intracluster medium (ICM) of approximately half
of all galaxy clusters is less than 1~Gyr, establishing a population of cool
core (CC) clusters \citep{Cavagnolo2009, Hudson2010}. Since this cooling time
falls below the cluster formation time by up to an order of magnitude, a copious
amount of cold gas is expected to precipitate from the hot gaseous atmospheres
and to form stars at rates up to several hundred $\rmn{M}_\odot~\rmn{yr}^{-1}$
\cite[see][for a review]{Peterson2006}. The absence of radiative cooling and
star formation at the predicted high rates calls for a heating mechanism that
stabilizes the system. A promising framework is provided by energy feedback from
an active galactic nucleus (AGN) at the cluster centre that accretes cooling gas
and launches relativistic jets, which inflate radio lobes that are co-localized
with the cavities seen in the X-ray maps. As the energy is transferred to the
surrounding gas, this offsets radiative cooling until the heating reservoir is
exhausted and the cooling gas can fuel the central AGN again, thus establishing
a tightly self-regulated feedback loop. However, there has been little direct
evidence supporting the existence of this hypothetical feedback cycle. In this
paper, we will provide empirical evidence for such a self-regulating
heating/cooling cycle and present a theoretical model to explain the underlying
physics.

\C{Because the energetics of AGN feedback is more than sufficient to balance
  radiative cooling, it has been suggested that AGN feedback can transform CC
  into non-CC clusters \citep{Guo2009, Guo2010b}. However, correlating the
  cavity enthalpy with the central gas entropy demonstrates that CC clusters
  cannot be transformed into non-CC clusters on the buoyancy time-scale due to
  the weak coupling of the mechanical to internal energy of the cluster gas
  \citep{2012ApJ...752...24P}.}  This calls for a {\em process that operates on
  a slower time-scale than the sound crossing time}.  Several physical processes
associated with the rising radio lobes have been proposed to be responsible for
the heating, including \C{mixing \citep{Kim2003Turb, Yang2016b}}, redistribution
of heat by buoyancy-induced turbulent convection \citep{2007ApJ...671.1413C,
  2009ApJ...699..348S} and dissipation of mechanical heating by outflows,
lobes or sound waves from the central AGN \citep[e.g.,][]{2001ApJ...554..261C,
  Brueggen2002, 2002ApJ...581..223R, 2012MNRAS.424..190G}.  \C{Also the
  role of thermal conduction in combination with AGN feedback has been explored
  \citep{Kannan2016, Yang2016a}.}

As those jet-inflated lobes rise in the cluster potential, they excite gravity
modes \citep{Reynolds2015}, which successively decay and generate turbulence that
dissipates and heats the cluster gas \citep[e.g.,][]{Zhuravleva2014}. Recent
micro-calorimetric X-ray observations of the core of the Perseus cluster find a
low ratio of the turbulent-to-thermal pressure of 4\% \citep{Hitomi2016}.  Such
low-velocity turbulence cannot propagate far from the excitation site without
being replenished, requiring turbulence to be generated in situ throughout the
core or to be transported (non-thermally) from the radio lobes.

There is an alternative that explains the slow dissipation rate (acting on the
Alfv{\'e}n crossing time) and operates homogeneously throughout the cluster
core. Relativistic particles (called cosmic rays, CRs) that are accelerated in
the relativistic jet are likely mixed into the ambient thermal plasma during the
buoyant rise of radio lobes \C{\citep{Sijacki2008, Guo2011, Pfrommer2013}}. As
they propagate from the injection site, they are following the ubiquitous
magnetic fields \citep{Kuchar2011} that redistribute their momenta to
homogeneously fill the central core before they propagate towards larger
radii. Fast-streaming CRs along the magnetic field excite Alfv{\'e}n waves
through the ``streaming instability'' \citep{Kulsrud1969}.  Scattering on this
wave field limits the macroscopic speed of GeV CRs to velocities of order the
Alfv{\'e}n speed.  Non-linear Landau damping of these Alfv{\'e}n waves provides
a means of transferring CR energy to the cooling gas.  This may provide an
efficient mechanism of suppressing the cooling catastrophe of cooling cores
\citep{Loewenstein1991, Guo2008, Ensslin2011, Fujita2011,Pfrommer2013, Wiener2013}.

To scrutinize this model, we compiled a large sample of 39 CC clusters in our
first companion paper \citep[][hereafter JP17]{Jacob2016a} and \C{found
steady-state solutions} that match all observed density and temperature profiles
well. In those models radiative cooling is balanced by CR heating in the cluster
centres and by thermal conduction on larger scales. Most importantly we found a
continuous sequence of cooling properties in our sample: clusters hosting radio
mini halos (RMHs) are characterized by the largest cooling radii, star formation
and mass deposition rates and thus signal the presence of a higher
cooling activity. Correspondingly, more CRs are needed to balance cooling in
those clusters.

RMHs are radio-emitting diffuse sources with typical radial extensions of 100 to
200~kpc that are centred on some CC clusters \citep{Giacintucci2014}.  The
detection of unpolarized radio synchrotron emission from RMHs proves the
existence of volume-filling magnetic fields and CR electrons in the cooling
regions of those clusters. In contrast, Mpc-sized giant radio halos occur in a
fraction of X-ray luminous non-CC clusters that are currently merging with
another cluster \citep[see e.g.][for a review]{Feretti2012}.

In this second paper about CR heating in CC clusters, we assess the
viability of our steady-state solutions by comparing the resulting non-thermal
radio and gamma-ray emission to observational data \C{ \citep[similarly
    to][]{Pfrommer2004,Colafrancesco2008,Fujita2012,Fujita2013}}. As CR protons
interact inelastically with the ambient gas protons, they produce primarily
pions (provided their energy exceeds the kinematic threshold of the
reaction). Neutral pions decay into gamma-rays and charged pions produce
secondary positrons and electrons that emit radio-synchrotron
radiation\footnote{\C{Throughout the paper, the term secondary electrons also
    includes secondary positrons.}}. Confronting our model predictions with data
enables us to put forward an observationally supported model for self-regulated
feedback heating, in which an individual cluster is either stably heated,
predominantly cooling, or is transitioning from one state to the other.

Our paper is structured as follows.  In Section~\ref{sec:properties}, we briefly
discuss the cluster sample and the density and CR pressure profiles that we base
our analysis on. In Sections~\ref{sec:radio} and \ref{sec:gamma}, we compare the
non-thermal emission of our steady-state solutions to observational radio and
gamma-ray data, respectively. In Section~\ref{sec:picture}, we present the emerging
picture of the self-regulation cycle of CC clusters and conclude in
Section~\ref{sec:conclusions}.

Throughout this paper, we use a standard cosmology with a present-day Hubble
factor $H_0=70~\rmn{km~s}^{-1}\rmn{Mpc}^{-1}$, and density parameters of matter,
$\Omega_{\rmn{m}} = 0.3$, and due to a cosmological constant, $\Omega_{\Lambda}
= 0.7$.

%%%%%%%%%%%%%%%%%%%%%%%%%%%%%%%%%%%%
\section{Cluster Properties}
\label{sec:properties}

Our analysis is based on the cluster sample in \citetalias{Jacob2016a}. Here,
we briefly introduce and characterize this sample, describe fits
to the density profiles and show the employed CR pressure profiles.

\subsection{Sample selection}

\label{sec:sample}

Our cluster sample consists of 39 CC clusters from the Archive of Chandra
Cluster Entropy Profile Tables, ACCEPT \citep{Cavagnolo2009} and has been
detailed in \citetalias{Jacob2016a}; here we only provide an overview of the
selection criteria. The clusters are chosen such that they either show
non-thermal emission or are promising targets for such an emission component. In
particular, the sample includes the 15 clusters with a confirmed RMH in
\citet{Giacintucci2014}. Most of the remaining clusters are selected due to the
high expected gamma-ray emission from pion decay \citep{Pinzke2011}. Since these
predictions derive from observed density profiles and a universal,
simulation-based CR model \citep{Pinzke2010}, they also represent the X-ray
brightest CCs for which {\em Chandra} data is available in the ACCEPT data
base.  Table~\ref{tab:sample} lists the clusters in our sample together with the
observed quantities from the literature that are relevant for our analysis.

The redshift distribution of this sample is not homogeneous since clusters with
RMHs have larger redshifts than most clusters without an RMH (shown in fig.~1
in \citetalias{Jacob2016a}). This is likely due to a selection bias that results
from an inherent surface brightness limit of a typical RMH observation, which
favours more compact, massive clusters at intermediate redshifts.  However, most
clusters in our sample have similar masses (within a factor of four),
independent of whether they host an RMH or not, and we have only a few outliers
at the low- and high-mass ends. To highlight the unbiased sample, we analyse our
core sample that is almost unbiased in mass and display the outliers for visual
purposes in more transparent colours where appropriate.

\begin{figure}
  \includegraphics{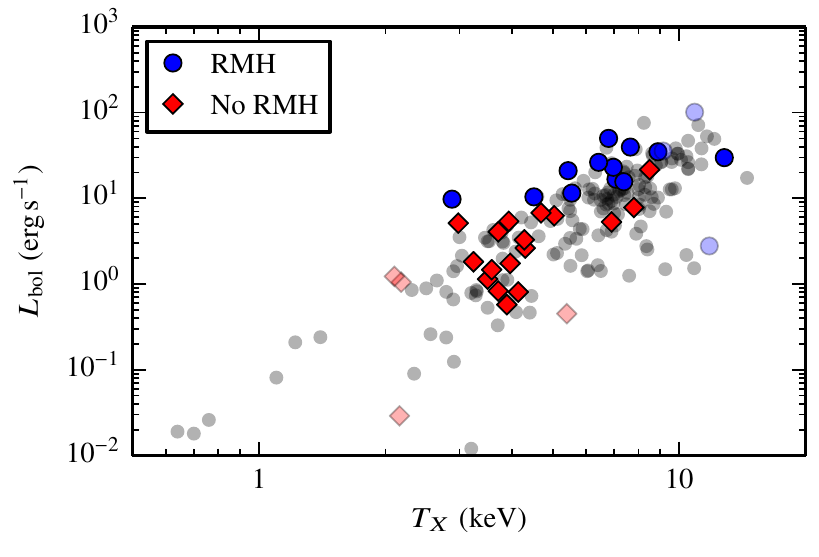}
  \caption{Bolometric X-ray luminosity and X-ray temperature (as an
    observational proxy for cluster mass) of all clusters in the ACCEPT data
    base (grey data points). Clusters of our sample are highlighted with blue
    circles if they host an RMH and with red diamonds if not. Remarkably,
    clusters with an RMH have typically higher bolometric luminosities than
    clusters without RMHs. Clusters at the low- and high-mass end (that do not
    belong to our core sample) are shown with transparent colours.}
    \label{fig:LbolvsTX}
\end{figure}

We further characterize our sample in Fig.~\ref{fig:LbolvsTX} by showing the
bolometric X-ray luminosity of all ACCEPT clusters as a function of the X-ray
temperature (an observational proxy for cluster mass). We highlight the CC
clusters of our sample with RMHs (blue circles) and the clusters without RMHs
(red diamonds). More transparent colours indicate our low- or high-mass
clusters, respectively. The figure shows that the selected clusters span the
whole parameter range that is covered by the ACCEPT sample. Still, clusters with
an RMH have systematically higher bolometric luminosities than clusters without RMHs.

While our unbiased cluster sample (full colours) spans a narrow range in
$M_{200}$ and $T_{\rmn{X}}$ of a factor of three, the bolometric X-ray
luminosity varies by over two orders of magnitudes, indicating the enormous
variance in core densities. CC clusters (including our entire sample) populate
the upper envelope of the $L_{\rmn{bol}}$--$T_{\rmn{X}}$ relation due to the
higher than average density of these systems at fixed cluster mass.

\begin{table*}
\caption{Cluster sample.}
\label{tab:sample}
\begin{threeparttable}
\begin{tabular}{l  r r r r r r r r r r r r r}
\hline
	 Cluster			&\multicolumn{1}{c}{$z$$^{\rmn{(1)}}$} 	&\multicolumn{1}{c}{$M_{200}^{\rmn{(2)}}$}	&\multicolumn{1}{c}{$kT_{\rmn{X}}^{\rmn{(1)}}$}&\multicolumn{1}{c}{SFR$_{\rmn{IR}}^\rmn{(3)}$}	 &\multicolumn{1}{c}{$r_\rmn{cool}^\rmn{(4)}$}		&\multicolumn{1}{c}{$F_{\nu,\rmn{NVSS}}^\rmn{(5)}$}	&\multicolumn{1}{c}{$F_{\nu, \rmn{mod}}$} &\multicolumn{1}{c}{$F_{\gamma,\rmn{obs}}^\rmn{(6)}$}& \multicolumn{1}{c}{$F_{\gamma, \rmn{mod}}$}& \multicolumn{1}{c}{$F_{\gamma,>1\rmn{~GeV}}$}\\
& & $\left(\times10^{14}\right)$& & & &&&$\left(\times 10^{-11}\right)$&$\left(\times 10^{-11}\right)$&$\left(\times 10^{-11}\right)$	 \\
 &	&\multicolumn{1}{c}{$\rmn{~M_\odot}$}	&  \multicolumn{1}{c}{keV} & \multicolumn{1}{c}{$\rmn{M_\odot\;yr^{-1}}$}&\multicolumn{1}{c}{kpc}&\multicolumn{1}{c}{mJy} & \multicolumn{1}{c}{mJy}&\multicolumn{1}{c}{$\rmn{ph\,cm^{-2}\,s^{-1}\!\!\!}$} &\multicolumn{1}{c}{$\rmn{ph\,cm^{-2}\,s^{-1}\!\!\!}$}&\multicolumn{1}{c}{$\rmn{ph\,cm^{-2}\,s^{-1}}\!\!\!$}  \\\hline
A 3112				&	0.0720	&	6.5$^{\rmn{a}}$		&	4.28	&	4.2$^{\rmn{a}}$\hphantom{0}	&	19.8	&	\dots\hphantom{000}					&	9.16$\times$10$^\rmn{1}$				&	27.3$^\rmn{a}$\hphantom{00}	&	1.75\hphantom{0}	&	1.75	\\
MKW3S				&	0.0450	&	5.1$^{\rmn{a}}$		&	3.50	&	\dots\hphantom{0}				&	6.6		&	$^\rmn{a\,}$1.15$\times$10$^\rmn{5}$	&	8.79\hphantom{$\times$10$^\rmn{5}$}	&	\dots\hphantom{.00}			&	\dots\hphantom{.0}	&	0.50	\\
Virgo (M87)			&	0.0044	&	1.4$^{\rmn{b}}$		&	2.50	&	0.24$^{\rmn{b}}$				&	9.5		&	$^\rmn{a\,}$1.39$\times$10$^\rmn{5}$	&	1.25$\times$10$^\rmn{1}$				&	135$^\rmn{b}$\hphantom{.000}	&	51.52\hphantom{0}	&	51.52	\\
A 2052				&	0.0353	&	4.4$^{\rmn{a}}$		&	2.98	&	1.4$^{\rmn{a}}$\hphantom{0}	&	15.0	&	$^\rmn{a\,}$5.50$\times$10$^\rmn{3}$	&	8.43\hphantom{$\times$10$^\rmn{5}$}	&	24$^\rmn{c}$\hphantom{.000}	&	2.88\hphantom{0}	&	0.80	\\
Centaurus			&	0.0109	&	5.3$^{\rmn{a}}$		&	3.96	&	0.18$^{\rmn{b}}$				&	10.9	&	$^\rmn{a\,}$3.80$\times$10$^\rmn{3}$	&	7.11\hphantom{$\times$10$^\rmn{5}$}	&	801$^\rmn{d}$\hphantom{.000}&	25.30\hphantom{0}	&	5.12	\\
Hydra A				&	0.0549	&	6.2$^{\rmn{a}}$		&	4.30	&	3.77$^{\rmn{b}}$				&	18.9	&	$^\rmn{a\,}$4.08$\times$10$^\rmn{4}$	&	1.00$\times$10$^\rmn{2}$				&	19.6$^\rmn{a}$\hphantom{00}	&	3.17\hphantom{0}	&	3.17	\\
A 4059				&	0.0475	&	6.6$^{\rmn{a}}$		&	4.69	&	0.57$^{\rmn{b}}$				&	7.3		&	$^\rmn{a\,}$1.28$\times$10$^\rmn{3}$	&	3.99\hphantom{$\times$10$^\rmn{5}$}	&	9.1$^\rmn{a}$\hphantom{00}	&	0.25\hphantom{0}	&	0.25	\\
A 262				&	0.0164	&	1.9$^{\rmn{a}}$		&	2.18	&	0.54$^{\rmn{a}}$				&	5.8		&	$^\rmn{a\,}$6.57$\times$10$^\rmn{1}$	&	0.22\hphantom{$\times$10$^\rmn{5}$}	&	9.3$^\rmn{a}$\hphantom{00}	&	0.13\hphantom{0}	&	0.13	\\
A 3581				&	0.0218	&	1.8$^{\rmn{a}}$		&	2.10	&	\dots\hphantom{0}				&	12.9	&	$^\rmn{b\,}$6.46$\times$10$^\rmn{2}$	&	2.27\hphantom{$\times$10$^\rmn{5}$}	&	110$^\rmn{c}$\hphantom{.000}	&	1.69\hphantom{0}	&	0.47	\\
A 2199				&	0.0300	&	6.4$^{\rmn{a}}$		&	4.14	&	0.58$^{\rmn{b}}$				&	13.1	&	$^\rmn{a\,}$3.58$\times$10$^\rmn{3}$	&	3.47$\times$10$^\rmn{1}$				&	19.8$^\rmn{a}$\hphantom{00}	&	4.74\hphantom{0}	&	4.74	\\
A 1644				&	0.0471	&	10.0$^{\rmn{a}}$	&	4.60	&	\dots\hphantom{0}				&	9.1		&	$^\rmn{b\,}$9.84$\times$10$^\rmn{1}$	&	1.07\hphantom{$\times$10$^\rmn{5}$}	&	16$^\rmn{a}$\hphantom{.000}	&	0.10\hphantom{0}	&	0.10	\\
MKW 4				&	0.0198	&	1.4$^{\rmn{a}}$		&	2.16	&	0.03$^{\rmn{b}}$				&	7.6		&	$^\rmn{b\,}$1.71$\times$10$^\rmn{1}$	&	0.36\hphantom{$\times$10$^\rmn{5}$}	&	\dots\hphantom{.00}			&	\dots\hphantom{.0}	&	0.13	\\
A 539				&	0.0288	&	4.4$^{\rmn{a}}$		&	3.24	&	\dots\hphantom{0}				&	2.5		&	$^\rmn{b\,}$6.3\hphantom{0$\times$10$^\rmn{3}$}	&	0.14\hphantom{$\times$10$^\rmn{5}$}	&	\dots\hphantom{.00}&	\dots\hphantom{.0}	&	0.05	\\
A 1795				&	0.0625	&	12.8$^{\rmn{a}}$	&	7.80	&	\dots\hphantom{0}				&	20.0	&	$^\rmn{a\,}$9.25$\times$10$^\rmn{2}$	&	1.23$\times$10$^\rmn{2}$				&	5.8$^\rmn{a}$\hphantom{00}	&	3.33\hphantom{0}	&	3.33	\\
A 2597				&	0.0854	&	5.7$^{\rmn{a}}$		&	3.58	&	3.23$^{\rmn{b}}$				&	34.1	&	$^\rmn{a\,}$1.88$\times$10$^\rmn{3}$	&	2.59$\times$10$^\rmn{2}$				&	4.4\hphantom{00}				&	3.33\hphantom{0}	&	3.33	\\
A 133				&	0.0558	&	6.5$^{\rmn{a}}$		&	3.71	&	\dots\hphantom{0}				&	18.7	&	$^\rmn{a\,}$1.67$\times$10$^\rmn{2}$	&	2.78$\times$10$^\rmn{1}$				&	7.6$^\rmn{a}$\hphantom{00}	&	0.96\hphantom{0}	&	0.96	\\
A 496				&	0.0328	&	7.1$^{\rmn{a}}$		&	3.89	&	\dots\hphantom{0}				&	17.0	&	$^\rmn{b\,}$1.21$\times$10$^\rmn{2}$	&	2.05$\times$10$^\rmn{1}$				&	25.2$^\rmn{a}$\hphantom{00}	&	1.45\hphantom{0}	&	1.45	\\
A 907				&	0.1527	&	6.4$^{\rmn{c}}$		&	5.04	&	\dots\hphantom{0}				&	8.5		&	$^\rmn{b\,}$6.86$\times$10$^\rmn{1}$	&	2.48$\times$10$^\rmn{1}$				&	\dots\hphantom{.00}			&	\dots\hphantom{.0}	&	0.22	\\
PKS 0745			&	0.1028	&	9.8$^{\rmn{a}}$		&	8.50	&	17.2$^{\rmn{a}}$\hphantom{0}	&	44.5	&	$^\rmn{b\,}$2.37$\times$10$^\rmn{3}$	&	1.37$\times$10$^\rmn{3}$				&	82$^\rmn{c}$\hphantom{.000}	&	45.26\hphantom{0}	&	12.65	\\
AWM 7				&	0.0172	&	7.2$^{\rmn{a}}$		&	3.71	&	\dots\hphantom{0}				&	5.4		&	$^\rmn{b\,}$2.9\hphantom{0$\times$10$^\rmn{3}$}	&	1.73\hphantom{$\times$10$^\rmn{5}$}	&	384$^\rmn{d}$\hphantom{.000}	&	4.73	&	0.96	\\
ZwCl 1742			&	0.0757	&	13.1$^{\rmn{a}}$	&	4.40	&	2.02$^{\rmn{b}}$				&	13.4	&	$^\rmn{b\,}$9.12$\times$10$^\rmn{1}$	&	5.82$\times$10$^\rmn{1}$				&	10.4$^\rmn{a}$\hphantom{00}	&	1.16\hphantom{0}	&	1.16	\\
A 1991				&	0.0587	&	0.9$^{\rmn{c}}$		&	5.40	&	\dots\hphantom{0}				&	17.8	&	$^\rmn{b\,}$3.90$\times$10$^\rmn{1}$	&	2.52$\times$10$^\rmn{1}$				&	\dots\hphantom{.00}			&	\dots\hphantom{.0}	&	0.83	\\	
A 383				&	0.1871	&	5.0$^{\rmn{c}}$		&	3.93	&	5.58$^{\rmn{b}}$				&	32.5	&	$^\rmn{c\,}$4.09$\times$10$^\rmn{1}$	&	9.79$\times$10$^\rmn{1}$				&	\dots\hphantom{.00}			&	\dots\hphantom{.0}	&	0.60	\\
A 85				&	0.0558	&	10.9$^{\rmn{a}}$	&	6.90	&	0.61$^{\rmn{b}}$				&	20.0	&	$^\rmn{b\,}$5.67$\times$10$^\rmn{1}$	&	1.91$\times$10$^\rmn{2}$				&	18$^\rmn{a}$\hphantom{.000}	&	5.91\hphantom{0}	&	5.91	\\
\hline
Perseus (A 426)		&	0.0179	&	8.6$^{\rmn{a}}$		&	6.79	&	34.46$^{\rmn{b}}$				&	34.2	&	$^\rmn{a\,}$2.28$\times$10$^\rmn{4}$	&	1.10$\times$10$^\rmn{2}$				&	0.014$^\rmn{e}$				&	0.003				&	38.27	\\
A 2029				&	0.0765	&	12.9$^{\rmn{a}}$	&	7.38	&	\dots\hphantom{0}				&	24.5	&	$^\rmn{b\,}$5.28$\times$10$^\rmn{2}$	&	2.56$\times$10$^\rmn{2}$				&	328$^\rmn{d}$\hphantom{.000}	&	23.14\hphantom{0}	&	4.68	\\
A 2390				&	0.2301	&	24.8$^{\rmn{d}}$	&	9.16	&	40.6$^{\rmn{b}}$\hphantom{0}	&	18.9	&	$^\rmn{b\,}$2.35$\times$10$^\rmn{2}$	&	2.37$\times$10$^\rmn{2}$				&	43$^\rmn{c}$\hphantom{.000}	&	4.73\hphantom{0}	&	1.32	\\
A 478				&	0.0883	&	11.7$^{\rmn{a}}$	&	7.07	&	2.39$^{\rmn{b}}$				&	32.0	&	$^\rmn{a\,}$3.69$\times$10$^\rmn{1}$	&	2.15$\times$10$^\rmn{2}$				&	12.7$^\rmn{a}$\hphantom{00}	&	2.63\hphantom{0}	&	2.63	\\
2A 0335+096		&	0.0347	&	4.5$^{\rmn{a}}$		&	2.88	&	7$^{\rmn{c}}$\hphantom{.00}	&	31.4	&	$^\rmn{a\,}$3.67$\times$10$^\rmn{1}$	&	3.31$\times$10$^\rmn{2}$				&	6.7$^\rmn{a}$\hphantom{00}	&	15.53\hphantom{0}	&	15.53	\\
A 2204				&	0.1524	&	8.3$^{\rmn{a}}$		&	6.97	&	14.7$^{\rmn{a}}$\hphantom{0}	&	41.1	&	$^\rmn{c\,}$6.93$\times$10$^\rmn{1}$	&	7.51$\times$10$^\rmn{2}$				&	13$^\rmn{c}$\hphantom{.000}	&	15.53\hphantom{0}	&	4.34	\\
Ophiuchus			&	0.0280	&	40.5$^{\rmn{a}}$	&	11.79	&	\dots\hphantom{0}				&	13.3	&	$^\rmn{b\,}$2.88$\times$10$^\rmn{1}$	&	3.51$\times$10$^\rmn{2}$				&	2622$^\rmn{d}$\hphantom{.000}	&	336.41\hphantom{0}&	68.01	\\
ZwCl 3146			&	0.2900	&	12.5$^{\rmn{c}}$	&	12.80	&	65.51$^{\rmn{b}}$				&	43.8	&	$^\rmn{d\,}$9.58$\times$10$^\rmn{1}$	&	1.21$\times$10$^\rmn{3}$				&	\dots\hphantom{.00}			&	\dots\hphantom{.0}	&	3.85	\\
MS 1455.0+2232	&	0.2590	&	4.8$^{\rmn{c}}$		&	4.51	&	9.46$^{\rmn{b}}$				&	44.6	&	$^\rmn{d\,}$1.93$\times$10$^\rmn{1}$	&	2.62$\times$10$^\rmn{2}$				&	\dots\hphantom{.00}			&	\dots\hphantom{.0}	&	0.97	\\
RX J1720.1+2638	&	0.1640	&	8.7$^{\rmn{c}}$		&	5.55	&	\dots\hphantom{0}				&	32.5	&	$^\rmn{d\,}$8.77$\times$10$^\rmn{1}$	&	1.35$\times$10$^\rmn{3}$				&	\dots\hphantom{.00}			&	\dots\hphantom{.0}	&	9.42	\\
A 1835				&	0.2532	&	17.5$^{\rmn{d}}$	&	7.65	&	235.37$^{\rmn{b}}$				&	49.2	&	$^\rmn{c\,}$3.93$\times$10$^\rmn{1}$	&	1.12$\times$10$^\rmn{3}$				&	\dots\hphantom{.00}			&	\dots\hphantom{.0}	&	5.02	\\
RX J1532.9+3021	&	0.3450	&	7.9$^{\rmn{c}}$		&	5.44	&	97$^{\rmn{a}}$\hphantom{.00}	&	51.1	&	$^\rmn{c\,}$2.28$\times$10$^\rmn{1}$	&	7.99$\times$10$^\rmn{2}$				&	\dots\hphantom{.00}			&	\dots\hphantom{.0}	&	2.26	\\
RX J1504.1-0248	&	0.2150	&	15.1$^{\rmn{c}}$	&	8.90	&	140$^{\rmn{d}}$\hphantom{.00} &	57.0	&	$^\rmn{b\,}$6.05$\times$10$^\rmn{1}$	&	2.19$\times$10$^\rmn{3}$				&	96$^\rmn{c}$\hphantom{.000}	&	33.70\hphantom{0}	&	9.42	\\
RBS 0797			&	0.3540	&	9.7$^{\rmn{c}}$		&	6.43	&	\dots\hphantom{0}				&	51.5	&	$^\rmn{a\,}$2.17$\times$10$^\rmn{1}$	&	9.21$\times$10$^\rmn{2}$				&	\dots\hphantom{.00}			&	\dots\hphantom{.0}	&	2.26	\\
RX J1347.5-1145	&	0.4510	&	26.1$^{\rmn{c}}$	&	10.88	&	\dots\hphantom{0}				&	37.8	&	$^\rmn{d\,}$4.59$\times$10$^\rmn{1}$	&	3.00$\times$10$^\rmn{3}$				&	257$^\rmn{d}$\hphantom{.000}	&	24.90\hphantom{0}	&	5.03	\\
\hline
\end{tabular}
\begin{tablenotes}
\item (1) Taken from the ACCEPT homepage \citep{Cavagnolo2009}
\item (2) a) \citet{Pinzke2011} b) \citet{Urban2011} c) $M_{500}$ from \citet{Lagana2013} d) $M_{500}$ from \citet{Ettori2010}, for c) and d) we use $M_{200} = 200 \times 4 \pi \rho_{\rmn{crit}} r_{200}^3/3$
\item (3) a) \citet{ODea2008} b) \citet{Hoffer2012} c) \citet{Donahue2007} d) \citet{Ogrean2010}
\item (4) We define the cooling radius $r_\rmn{cool}$ as the radius where the cooling time is $1\rmn{~Gyr}$.
\item (5) a) \citet{Birzan2004} b) sources from the NVSS Source catalogue browser with distance to ACCEPT coordinates $<15\rmn{~arcsec}$, except for A~539 ($1.08\rmn{~arcmin}$) and MKW~4 ($1.32\rmn{~arcmin}$)
c) \citet{Sayers2013} d) \citet{Coble2007}
\item (6) a) \citet{Fermi2014} ($>1\rmn{~GeV}$) b) \citet{Abdo2009M87} ($>1\rmn{~GeV}$) c) \citet{Dutson2013} ($>0.3\rmn{~GeV}$) d) \citet{Fermi2010} ($0.2 - 100\rmn{~GeV}$) e) \citet{Magic2012Perseus} ($>1\rmn{~TeV}$)\C{; these values are upper limits except for Virgo/M87 \citep{Abdo2009M87}}
\end{tablenotes}
\end{threeparttable}
\end{table*}

\subsection{Density profiles}
\label{sec:density}

The non-thermal radio and gamma-ray emission depend on the density profiles of
the clusters, either directly since the hadronic reaction is a two-body process
with an emissivity that scales with the product of gas and CR density or
indirectly through the magnetic field, which assumes a density dependence
through the magnetic flux-freezing condition. Here, we use fits to
observationally inferred density profiles.  If we were to use the density
profiles of our steady-state solutions as derived in \citetalias{Jacob2016a},
this would only result in small changes and have no influences on our
conclusions.

For 15 clusters, we use the fits by \citet{Vikhlinin2006} and \citet{Landry2013} who use the formula 
\begin{equation}
  n_\rmn{p} n_\rmn{e} =
  \frac{n_0^2\,\left( r/r_{\rmn{c}} \right)^{-\alpha}}
       {\left[ 1 + \left(r/r_{\rmn{c}}\right)^2 \right]^{3 \beta - \alpha/2}}
  \frac{1}{\left[1 + \left(r/r_{\rmn{s}}\right)^{\gamma} \right]^{\varepsilon/\gamma}}
+\frac{n_{\rmn{02}}^2}{\left[ 1 + \left(r/r_{\rmn{c2}}\right)^2\right]^{3 \beta_2}},
\end{equation}
where $n_\rmn{e}/n_{\rmn{p}} = 1.19$ (see also Appendix~\ref{sec:appradio}).

We obtain the density profiles for the remaining clusters by performing our own
fits. To this end, we use the data points provided on the ACCEPT homepage. Since
the \textit{Chandra} data only cover the centres of most clusters, we find that a single
beta profile is sufficient to describe the data and adopt the following profile
\begin{equation}
n_\rmn{e} = n_0 \left[ 1 + \left(r/r_{\rmn{c}} \right)^2\right]^{- 3 \beta/2}
\end{equation}
in a suitable radial range. The fit results together with the radial range of
applicability can be found in Appendix~\ref{sec:appdensity} in
Table~\ref{tab:fitparam}.

\subsection{CR pressure profiles}

We use a CR population that is able to balance radiative cooling in the centres
of CC clusters through the excitation of Alfv{\'e}n waves. These waves get
dissipated, which implies a volumetric heating rate \citep{Wentzel1971}
\begin{equation}
  \mathcal{H}_{\rmn{cr}} = - \bmath{\vv}_{\rmn{st}} \bmath{\cdot \nabla} P_{\rmn{cr}}.
\end{equation}
Here, $P_{\rmn{cr}}$ is the CR pressure and the streaming velocity is given by
\begin{equation}
   \bmath{\vv}_{\rmn{st}} =
   - \rmn{sgn}(\bmath{B \cdot \nabla} P_{\rmn{cr}}) \bmath{\vv}_{\rmn{A}},
\end{equation}
where $\bmath{B}$ is the magnetic field, $\bmath{\vv}_{\rmn{A}} = \bmath{B}/
\sqrt{4 \uppi \rho}$ is the Alfv{\'e}n velocity and $\rho$ is the mass
density. 

Such a CR population is described by the steady-state solutions from
\citetalias{Jacob2016a}, which are obtained by solving the hydrodynamic
equations that are coupled to the CR energy equation. In those solutions,
radiative cooling is balanced by thermal conduction at large scales and CR
heating in the central regions, giving rise to a small mass deposition rate of
cooling gas that precipitates out of the hot atmosphere at a level that is
typically 10 per cent of the observed infrared (IR) star formation rate (SFR).  The resulting CR
pressure profiles approximately follow the thermal pressure profiles in the
central region (i.e., $X_\CR\equiv P_\CR/P_\rmn{th}\approx\rmn{const.}$) and
fall off rapidly outside the radius at which conductive heating starts to
dominate.  The model parameters are chosen such that the solutions are physical
and that the radial extent over which CR heating dominates is maximized (for
more detail please refer to \citetalias{Jacob2016a}).

The steady-state solutions are only valid in a certain radial range, i.e.,
between the radii $r_\rmn{in}$ and $r_\rmn{out}$.  To determine the non-thermal
emission, we extend the solutions to the centre of the cluster with a constant
value. Since the outer radius can vary substantially from cluster to cluster, we
extrapolate the profile to $200\rmn{~kpc}$ if $r_\rmn{out}$ is smaller than
that, which is the case for 25 out of 39 clusters. Beyond $200\rmn{~kpc}$, the
CR pressure has typically dropped significantly, such that the hadronically
induced fluxes are fully determined by the emission at smaller radii and the
exact cut-off radius becomes unimportant. For the extrapolation to larger radii,
we use a power law. The final CR pressure profile is then given by
\begin{equation}
  \label{eq:Pex}
P_\rmn{cr, ex} (r) =
\begin{cases}
P_\rmn{cr}(r_\rmn{in}) 	&  r < r_\rmn{in}\\
P_\rmn{cr}(r) 		& r_\rmn{in} < r < r_\rmn{out}\\
P_\rmn{cr}(r_\rmn{out})\left(\frac{r}{r_\rmn{out}}\right)^{\alpha_{P_\rmn{cr}}}
                        &  r > r_\rmn{out}
\end{cases}
\end{equation}
with $\alpha_{P_\rmn{cr}} = \left.\rmn{d} \ln P_\rmn{cr}/\rmn{d} \ln r
\right|_{r_\rmn{out}}$.

%%%%%%%%%%%%%%%%%%%%%%%%%%%%%%%%%%%%
\section{Radio emission}
\label{sec:radio}

Hadronic interactions between relativistic CRs and the ambient cluster medium
lead to secondary electrons and hence synchrotron emission.  Here, we compare
the modelled radio emission of our steady-state CR population to observed data
by the NRAO VLA Sky Survey \citep[NVSS,][]{Condon1998} as well as RMHs.

\begin{figure*}
  \includegraphics{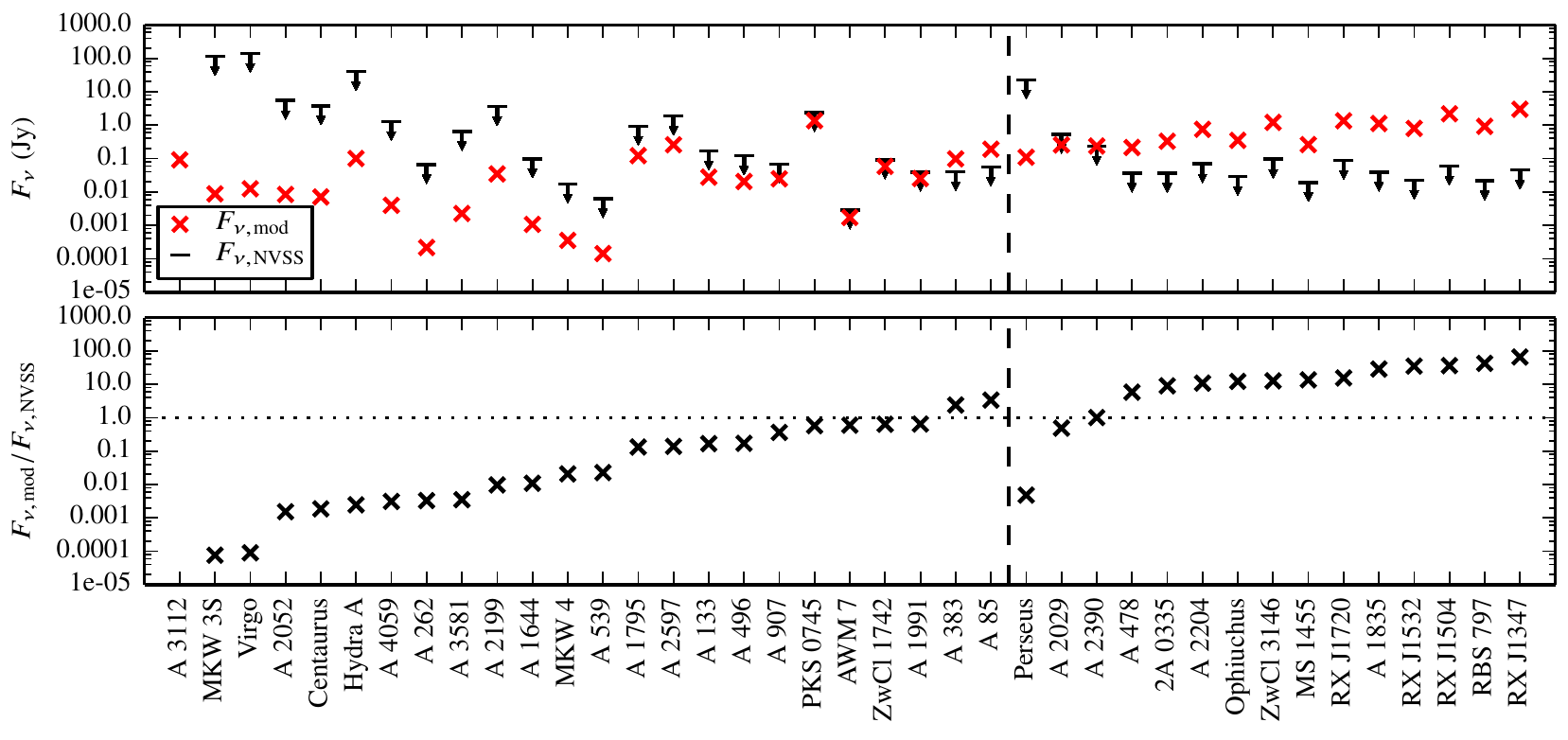}
  \caption{Comparison between the predicted secondary radio flux of our steady-state solutions and the 1.4 GHz flux measured by NVSS. Because the radio
    emission observed by NVSS likely acquires a partial contribution from
    primary accelerated CR electrons, it represents an upper limit to the
    hadronically generated secondary radio emission. The top panel shows the
    absolute flux values of a new predicted class of {\em radio micro halos}
    (left to the dashed line). In the bottom panel, we display the ratio of
    predicted to observed flux. For most clusters without RMHs the predicted
    flux is smaller than the flux observed by NVSS, whereas for RMH clusters the
    predicted flux is generally in conflict with the data. This excludes CR
    pressures at a level required to stably balance radiative cooling in most
    clusters hosting an RMH.}
  \label{fig:nvssplot}
\end{figure*}

%----------------------------------------------------------------------------------------------------
\subsection{Emissivity}
\label{sec:emissivity}

We calculate the radio emission closely following \citet{Pfrommer2008}. For
brevity, we only introduce the most important concepts here and provide the
complete description in Appendix~\ref{sec:appemission}.

The distribution of protons, which we describe in terms of the dimensionless
proton momentum $p_\rmn{p} = P_\rmn{p}/(m_\rmn{p} c)$, is given by
\begin{equation}
  f_\rmn{p} (p_\rmn{p}) = \frac{\rmn{d} N}{\rmn{d} p_\rmn{p} \rmn{d} V} =
  C_\rmn{p}(r) p_\rmn{p}^{-\alpha_\rmn{p}} \theta(p_\rmn{p}-q_\rmn{p}).
\end{equation}
It represents a power-law in momentum with spectral index
$\alpha_\rmn{p}=2.4$\footnote{\C{Note that our results are robust to changes in
    $\alpha_p$ by $\pm 0.3$.}}. Such a spectral index is expected for a CR
population that was injected by an AGN and may have experienced a mild spectral
steepening as a result of outwards streaming \citep{Wiener2013}.  We enforce a
lower momentum cut-off at $q_\rmn{p}=0.5$ with the Heaviside step function
$\theta(x)$. We denote the normalization by $C_\rmn{p}(r)$ and specify it with
the steady-state solutions for the CR pressure.

Hadronic CR proton interactions with the ICM produce secondary electrons.  If
radiative losses are taken into account, this population of secondary electrons
reaches a steady state with a spectral index that is steepened by one,
$\alpha_\rmn{e}=\alpha_\rmn{p}+1$ \citep{Sarazin1999}. The corresponding
secondary synchrotron emissivity $j_\nu$ at frequency $\nu$ per steradian is
given by
\begin{equation}
  \label{eq:jnu}
  j_\nu = \frac{A_\nu}{4 \upi} C_\rmn{p} n_\rmn{N}
  \frac{e_B}{e_B + e_\rmn{rad}}
  \left(\frac{e_B}{e_{B_\rmn{c}}}\right)^{(\alpha_\nu -1)/2}.
\end{equation}
The emissivity depends on the normalization of the CR protons $C_\rmn{p}$ and on
the nucleon density $n_\rmn{N}$, which is proportional to the electron number
density introduced in Section~\ref{sec:density}. \C{Moreover, it depends on the
  frequency-dependent normalization factor, $A_\nu$, the magnetic energy
  density, $e_B$, the energy density in radiation, $e_\rmn{rad}$, a frequency
  dependent characteristic magnetic field strength, $e_{B_\rmn{c}}$, and the
  radio spectral index $\alpha_\nu=(\alpha_\rmn{e}-1)/2$.}

Assuming an isotropic distribution of the CR electrons' pitch angles, the
synchrotron emissivity can be written in terms of the magnetic energy density
$e_B = B^2/(8 \upi)$. We parametrize the magnetic field strength as in
\citetalias{Jacob2016a}, which is motivated by analyses of deprojected Faraday
rotation measure maps and minimum field estimates by radio observations with the LOw Frequency ARray
\citep{Vogt2005, Kuchar2011, deGasperin2012},
\begin{equation}
B(r) = B_0 \left(\frac{n_\rmn{e}(r)}{0.01\rmn{~cm^{-3}}} \right)^{\alpha_B}.
\label{eq:Bfield}
\end{equation}
We adopt a magnetic field normalization $B_0=10\,\rmn{\umu G}$ and a power law
index of $\alpha_B = 0.5$.  This choice implies a radially constant Alfv{\'e}n
speed $\vv_\rmn{A}$.

Additionally, the energy density from radiation fields attains contributions
from the cosmic microwave background (CMB) and from stars and dust (SD) in the
central galaxy, $e_\rmn{rad}$. As a result, the synchrotron emissivity is
modified because these emitting CR electrons suffer additional energy losses
from inverse Compton scattering on the total radiation field. In the regime
of weak fields ($e_B \ll e_\rmn{rad}$), the emissivity strongly depends on the
magnetic field strength (see equation~\ref{eq:jnu}). The radiation from stars
and dust predominates within the central galaxy and exceeds the magnetic energy
density up to a radius of $20$ -- $40\rmn{~kpc}$, depending on the particular
cluster (see Fig.~\ref{fig:energydensities}). At larger radii, the magnetic
field starts to predominate as long as its energy density exceeds the energy
density of the cosmic microwave background, $\eps_B>\eps_\rmn{CMB}$, which is
the case for $n_\rmn{e}\gtrsim10^{-3}\,\rmn{cm}^{-3}$ or equivalently most of
the cool core regions studied in this work ($r\lesssim130 -- 200$~kpc according
to Fig.~\ref{fig:energydensities}).

The emissivity scales with frequency as $j_\nu\propto\nu^{-\alpha_\nu}$. Here,
$\alpha_\nu=(\alpha_\rmn{e}-1)/2=1.2$, which is encapsulated in $e_{B_\rmn{c}}$
and $A_\nu$. \C{In the limit of large magnetic fields ($e_B \gg e_\rmn{rad}$),
  the electrons loose all their energy to synchrotron radiation and not by
  inverse Compton scattering. Thus, to good approximation we can neglect the
  energy density in radiation in Equation~\eqref{eq:jnu}. In this case, we obtain
  a weak scaling of the emissivity with magnetic field strength since then
  $j_\nu\propto \eps_B^{(\alpha_\nu -1)/2}\approx \eps_B^{0.1}$ for our choice
  of $\alpha_\rmn{e}$.}  Because most of the secondary synchrotron emission is
collected from radii between 20 and 100~kpc for which we are clearly in the
magnetically dominated emission regime, the emissivity is mostly insensitive to
the exact value of magnetic field strength and is thus directly proportional to
the normalization of the CR distribution.

Using this emissivity, we can calculate theoretically expected surface brightness
profiles, luminosities and fluxes for the available observations. In the main
text, we focus on fluxes and surface brightness profiles, which we can directly
compare to observations, and discuss the luminosities in
Appendix~\ref{sec:applum}.

%----------------------------------------------------------------------------------------------------
\subsection{Comparison with NVSS data}
\label{sec:nvss}

We first compare the emission from the steady state CR population to the data
from the NVSS. This survey detects point sources at 1.4 GHz with a restored
beam of $45\rmn{~arcsec}$ full width half-maximum (FWHM). These data include
emission by primary electrons that are accelerated by the AGN combined with the
secondary electrons injected in hadronic interactions between CRs and thermal
protons. Therefore, the NVSS data have to be considered as upper limits for our
purpose.

We track the radial extent of the CR population to a maximum radius of
$r_{\rmn{max}, \parallel} = \max\left\{ r_\rmn{out}, 200\rmn{~kpc}\right\}$.
This choice ensures that we account for the entire CR energy in our non-thermal
emission because in most clusters, the CR pressure drops steeply at radii well
below $200\rmn{~kpc}$ as a result of CR streaming.  Moreover, this
characteristic radius corresponds to a typical radial extent of an RMH.  We
verified that the radio flux does not depend on the precise choice of this
radius because it is dominated by the central regions.  However, $r_{\rmn{max},
  \parallel}$ subtends an angle on the sky that is larger than the NVSS beam
width for all clusters.

Hence for the flux calculations, we first project the emissivity along the
radial direction and obtain the surface brightness as
\begin{equation}
S_\nu (r_\bot) = 2 \int_{r_\bot}^{r_\rmn{max, \parallel}} \rmn{d} r \frac{r j_\nu(r)}{\sqrt{r^2 - r_\bot^2}}.
\label{eq:snu}
\end{equation}
To determine the fluxes as seen by NVSS, we cut out a cylinder with radius
$r_{\rmn{max}, \bot}$ that corresponds to $22.5\rmn{~arcsec}$, half of the FWHM
of the beam, such that
\begin{equation}
F_\nu = 2 \upi \int_0^{r_{\rmn{max}, \bot}} \rmn{d} r_\bot r_\bot S_\nu \left( r_\bot \right).\label{eq:fluxfromsnu}
\end{equation}
We present the resulting fluxes together with the
observations in Fig.~\ref{fig:nvssplot} and list them in
Table~\ref{tab:sample}.\footnote{There is no data for A~3112 since its
  position on the sky was not observed by the NVSS.}

In the upper panel of Fig.~\ref{fig:nvssplot}, we show the absolute values of
the predicted flux at 1.4 GHz as well as the radio fluxes observed by NVSS. The
bottom panel shows their ratio. We separate clusters with and without an RMH and
order each group according to the flux ratio. The upper panel shows that the
predicted fluxes span orders of magnitude ranging from $10^{-4}$ to
$10\rmn{~Jy}$. The synchrotron flux predictions for clusters without an RMH are
significantly smaller than for clusters hosting an RMH, whereas the fluxes from
NVSS are often larger for clusters without RMHs.

There is an even stronger correlation in the flux ratios. Due to our ordering of
the clusters, the flux ratio increases from left to right. Interestingly, the
flux ratios for clusters with an RMH are generally much larger than for clusters
without an RMH. Moreover, there is a smooth transition from the clusters without
to the clusters with an RMH. An exception is Perseus with a very small
flux ratio. The reason for this is the exceptionally strong NVSS source Perseus
A since the predicted flux is in line with that of the other clusters.

We find that most flux ratios in clusters without an RMH are smaller than unity
but almost exclusively exceed unity in RMH clusters. {\em Thus, the level of CR
  pressure required to stably heat the interiors is in conflict with radio
  observations for RMH clusters while the secondary radio emission resulting
  from hadronic CR interactions is well below the observed fluxes in clusters
  without RMHs.} Together with the gradual transition between the two
populations this may indicate a self-regulated feedback loop. On the one side,
the cooling gas in non-RMH clusters may be stably balanced by CR heating, while
RMH clusters appear to be out of balance and predominantly cooling. This
interpretation is further discussed in Section~\ref{sec:picture}.

\begin{table}
\caption{Properties of the radio mini halos.}
\label{tab:rmhs}
\begin{threeparttable}
\begin{tabular}{l  r r r}
\hline
	 Cluster			&\multicolumn{1}{c}{$r_{\rmn{RMH}}^\rmn{(1)}$}&\multicolumn{1}{c}{$F_{\rmn{RMH, obs}}^\rmn{(1\C{,2})}$} & \multicolumn{1}{c}{$F_{\rmn{RMH, mod}}^\rmn{\C{(2)}}$} \\
& \multicolumn{1}{c}{kpc}&\multicolumn{1}{c}{mJy} &\multicolumn{1}{c}{mJy}  \\\hline
Perseus (A 426)		&	130	&	3020\hphantom{.0}	&	4914	\\
A 2029				&	270	&	19.5	&	728	\\
A 2390				&	250	&	28.3	&	348	\\
A 478				&	160	&	16.6	&	411	\\
2A 0335+096		&	70	&	21.1	&	1475	\\
A 2204				&	50	&	8.6		&	688	\\
Ophiuchus			&	250	&	83.4	&	8718	\\
ZwCl 3146			&	90	&	5.2		&	1184	\\
MS 1455.0+2232	&	120	&	8.5		&	288	\\
RX J1720.1+2638	&	140	&	72.0	&	1989	\\
A 1835				&	240	&	6.1		&	1449	\\
RX J1532.9+3021	&	100	&	7.5		&	782	\\
RX J1504.1-0248	&	140	&	20.0	&	2637	\\
RBS 0797			&	120	&	5.2		&	946	\\
RX J1347.5-1145	&	320	&	34.1	&	3221	\\
\hline
\end{tabular}
\begin{tablenotes}
\item (1) \citet{Giacintucci2014} and references therein
\item \C{(2) All fluxes correspond to $\nu = 1.4\rmn{~GHz}$.}
\end{tablenotes}
\end{threeparttable}
\end{table}

%----------------------------------------------------------------------------------------------------
\subsection{Radio mini halos}

The sources detected by the NVSS are point sources and can only be upper limits
for our predictions since they also include primary emission from the central
galaxy and its AGN.  The observed RMHs allow us to test the extended radio
emission from a CR population that is able to balance cooling.  To this end, we
study the fluxes of all RMHs and compare the surface brightness profiles of
individual clusters to the observations by \citet{Murgia2009}. In the end, we
discuss the robustness of our conclusions with respect to changes in the
parametrization of the magnetic field.

\subsubsection{RMH fluxes}

\begin{figure}
  \includegraphics{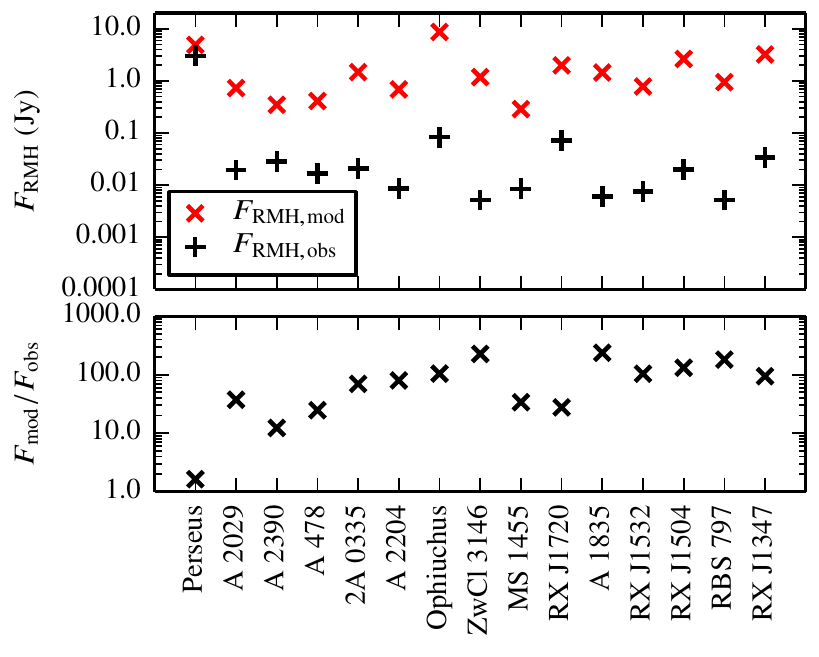}
  \caption{Comparison between the predicted fluxes for the RMHs and the
    observations from \citet{Giacintucci2014}. All predicted secondary radio
    fluxes exceed the observations by a substantial margin (with the exception
    of Perseus that is only barely excluded). Thus, this excludes CR pressures
    at a level that is required to stably balance radiative cooling in the
    central cluster regions exhibiting an RMH.}
    \label{fig:rmhfluxesplot}
\end{figure}

Here, we compare our modelled secondary RMH fluxes to the observed values \C{at $1.4\rmn{~GHz}$} in
\citet{Giacintucci2014}.  The hadronically induced RMH fluxes \C{at this frequency} from our CR
population are determined as in Section~\ref{sec:nvss}. In contrast to the
previous calculation, we now integrate the radio flux out to the radius
$r_{\rmn{max, \bot}} = \min\left\{r_\rmn{RMH},
r_{\rmn{max,\parallel}}\right\}$. The radius $r_\rmn{RMH}$ denotes the (average)
radius of the RMH as determined by \citet[see
Table~\ref{tab:rmhs}]{Giacintucci2014}.

We show the results in Fig.~\ref{fig:rmhfluxesplot}. The upper panel displays
the model predictions and observational fluxes, the lower panel their
ratio. Clearly, the predicted flux exceeds the observed flux in all clusters by
up to three orders of magnitude. This demonstrates that the secondary radio
emission from a CR population that is able to balance radiative cooling is
excluded by data. Conversely, this also means that if RMHs are powered by
hadronic CR interactions, those CRs have insufficient pressure to heat the
cluster gas.

While Perseus is formally excluded based on a moderate overproduction of the RMH
flux by a factor of 1.6, uncertainties in the magnetic field model and the
extent of the CR distribution along the line of sight could make it consistent
with the observational RMH data.

\subsubsection{Surface brightness profiles}

\begin{figure*}
  \includegraphics{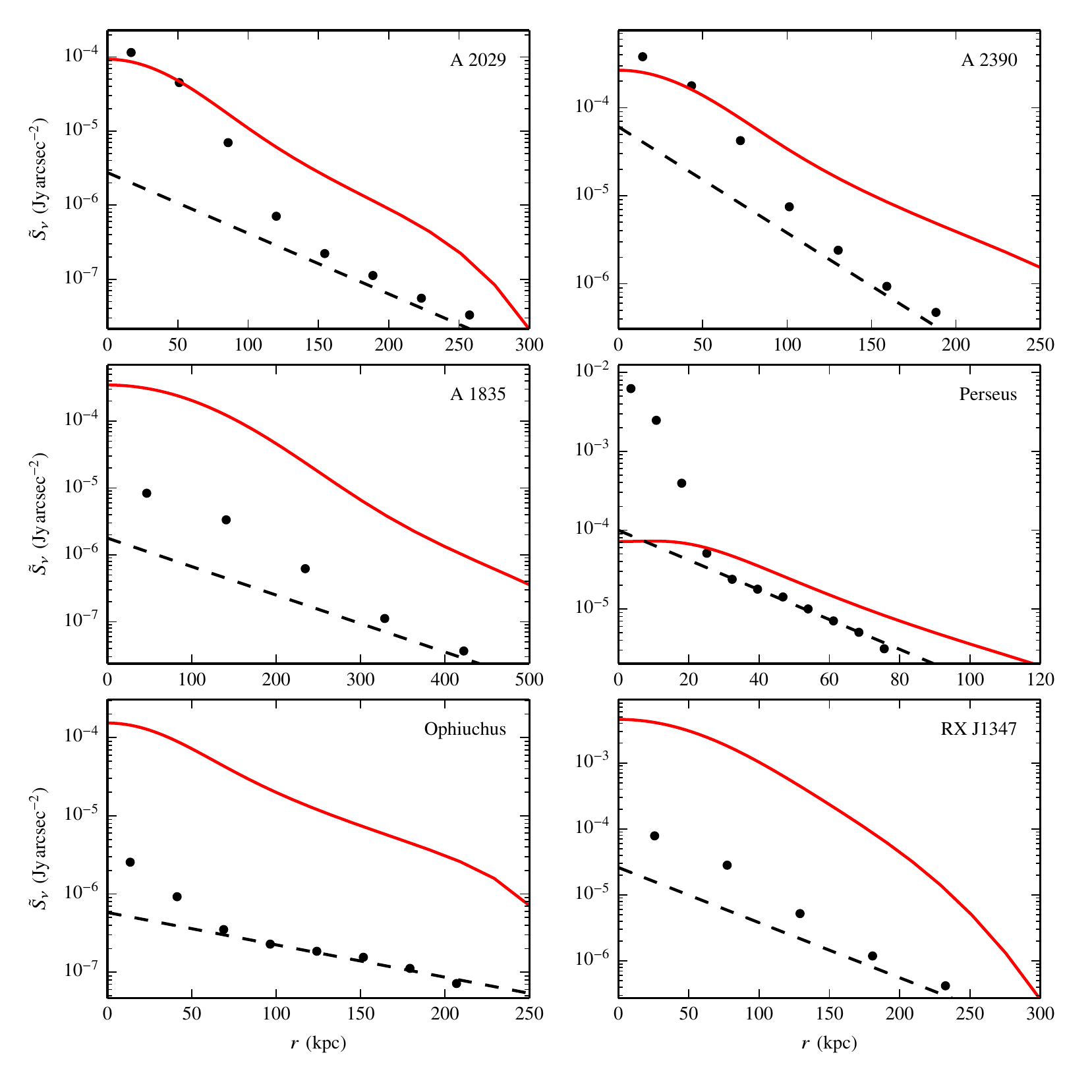}
  \caption{We compare the predicted radio surface brightness profiles of
    secondary synchrotron emission (red) to data from \citet{Murgia2009} (black
    data points). The black dashed lines show their fits to the emission
    from the RMH after modelling the central AGN emission. The expected emission
    exceeds the observed data by up to two orders of magnitudes, which indicates
    that CR heating is not viable in those clusters at all scales (with the
    exception of Perseus that is only marginally excluded).}
    \label{fig:radiodatacomp}
\end{figure*}

\citet{Murgia2009} analyse the surface brightness profiles of a sample of
clusters with radio halos and RMHs. Six of their clusters also coincide with
members of our sample so that we can test our model profiles.  The surface
brightness is in principle given by Equation~\eqref{eq:snu}, but for better
comparison we smooth our brightness profiles to the resolution of the Very Large
Array observations at 1.4 GHz as described in \citet{Murgia2009}.  Therefore, we
convolve the surface brightness with a Gaussian beam of standard deviation
$\sigma = \rmn{FWHM}_\rmn{beam}/(2 \sqrt{2 \log(2)})$,
\begin{eqnarray}
  \tilde{S}_\nu (r) &=&\!
  \frac{1}{2\upi\sigma^2}\int \rmn{d}^2 x' S_\nu (|\bmath{x'}|)
  \exp\left(- \frac{(\bmath{x}-\bmath{x'})^2}{2\sigma^2}\right)\\
  &=&\! \frac{1}{\sigma^2} \int_0^{\infty} \rmn{d}x x
  S_\nu (x) \exp\left(- \frac{x^2 + r^2}{2 \sigma^2}\right)
  I_0 \left( \frac{r x}{\sigma^2}\right),
  \nonumber
\end{eqnarray}
where $I_0(x)$ denotes a modified Bessel function of the first kind. For the
convolution, we assume that the surface brightness profile has dropped to zero
beyond $r_\rmn{max, \parallel}$.

In Fig.~\ref{fig:radiodatacomp}, we compare the expected surface brightness
profiles (red) to the radio data (black dots) of \citet{Murgia2009}. These data
contain the central radio source and the RMH. After modelling the central AGN,
the RMH contribution is shown as black dashed lines, which take the form of
exponential profiles \citep{Murgia2009}. The modelled secondary profiles exceed
the observed RMH profiles by a factor of two in Perseus and up to two orders of
magnitude in RX~J1347. In three cases the profiles exceed even the emission from
the central galaxy.  This demonstrates that the emission from a CR population
that is able to balance radiative cooling would overproduce the radio emission
in the core region delineated by the RMH emission, at least in those six
clusters.

Hence, our predictions generally surpass the limits set by radio observations in
clusters hosting an RMH, irrespective of whether we use NVSS data, RMH fluxes or
surface brightness profiles. Perseus and A2390 are only excluded by a factor of
a few to several and represent thus transitional objects. These systems can be
made consistent with the observational radio data by either lowering $X_\CR$ and
increasing $B_0$ by the same factor or by truncating the CR distribution along
the line of sight, which would lower the predicted radio flux without affecting
the central heating rate.  With the exception of those clusters, CR heating
plays no central role in balancing radiative cooling in RMH-hosting clusters.
Before we turn our attention to the gamma-ray emission, we assess the robustness
of our conclusions when varying the magnetic field model.

\subsubsection{Modifying the magnetic field}

Aside from the CR population, the CR heating rate and the radio emissivity
depend on the magnetic field strength. In our model, we fix the normalization at
$B_0 = 10\rmn{~\umu G}$ at $n_\rmn{e}=10^{-2}\, \rmn{cm}^{-3}$ (see
Section~\ref{sec:emissivity}). Here, we investigate whether it is possible to find a
combination of magnetic field and CR pressure that reproduces the RMH fluxes
and still balances radiative cooling.

In the limit of strong magnetic fields ($\eps_B\gg \eps_{\rmn{CMB}}$), the
emissivity is proportional to $j_\nu \propto C_\rmn{p} B^{\alpha_\nu -1}$. For
our choices of the spectral index, $\alpha_\nu = 1.2$ (see
Appendix~\ref{sec:appradio}), which means that the dependence of $j_\nu$ on the
magnetic field is extremely weak. Hence, the emissivity and therefore surface
brightness profiles and fluxes depend almost entirely on the CR population. In
order to meet the fluxes from the RMH, we would need to reduce the number of CRs
by at least a factor of 10 for most clusters (barring Perseus).  If the shape
of the CR profile remains the same, the CR heating rate is proportional to
$\mathcal{H}_\rmn{cr} \propto B C_\rmn{p}$. To achieve the same amount of CR
heating with the reduced CR population, we would thus have to increase the
magnetic field by a factor of 10. However, magnetic fields of $B_0 \approx
100\rmn{~\umu G}$ or higher would imply a plasma $\beta$ factor (i.e., the ratio
of thermal-to-magnetic pressure) of 0.1 instead of the observed values that are
of order or larger than 20 in cool core regions. Such a strong magnetic field is
excluded by Faraday rotation measurements and minimum energy arguments
\citep{Vogt2005, Kuchar2011, deGasperin2012} and would be impossible to grow and
maintain with a (turbulent) magnetic dynamo in the presence of a small
turbulent-to-thermal energy density ratio of 4 per cent \citep{Hitomi2016}.

For that reason, it is not possible to simultaneously reproduce the RMH fluxes
and heat the cluster gas with CRs {\em in the entire cool core region.} One
resort would be to refrain from maximal CR models that heat the entire radio
emitting region of RMHs. Instead, we could concentrate on CR heating models for
the central region that would dramatically reduce the required amount of CR
energy and by extension also the level of secondary radio emission to get into
agreement with RMH data. However, as we will discuss in
Section~\ref{sec:picture}, we present an alternative scenario that argues for
a heating/cooling imbalance in RMH clusters, which show strong signs of cooling
and star formation and for a stable balance in clusters without an observable
RMH.

%%%%%%%%%%%%%%%%%%%%%%%%%%%%%%%%%%%%

\section{Gamma-ray emission}
\label{sec:gamma}

\begin{figure*}
  \includegraphics{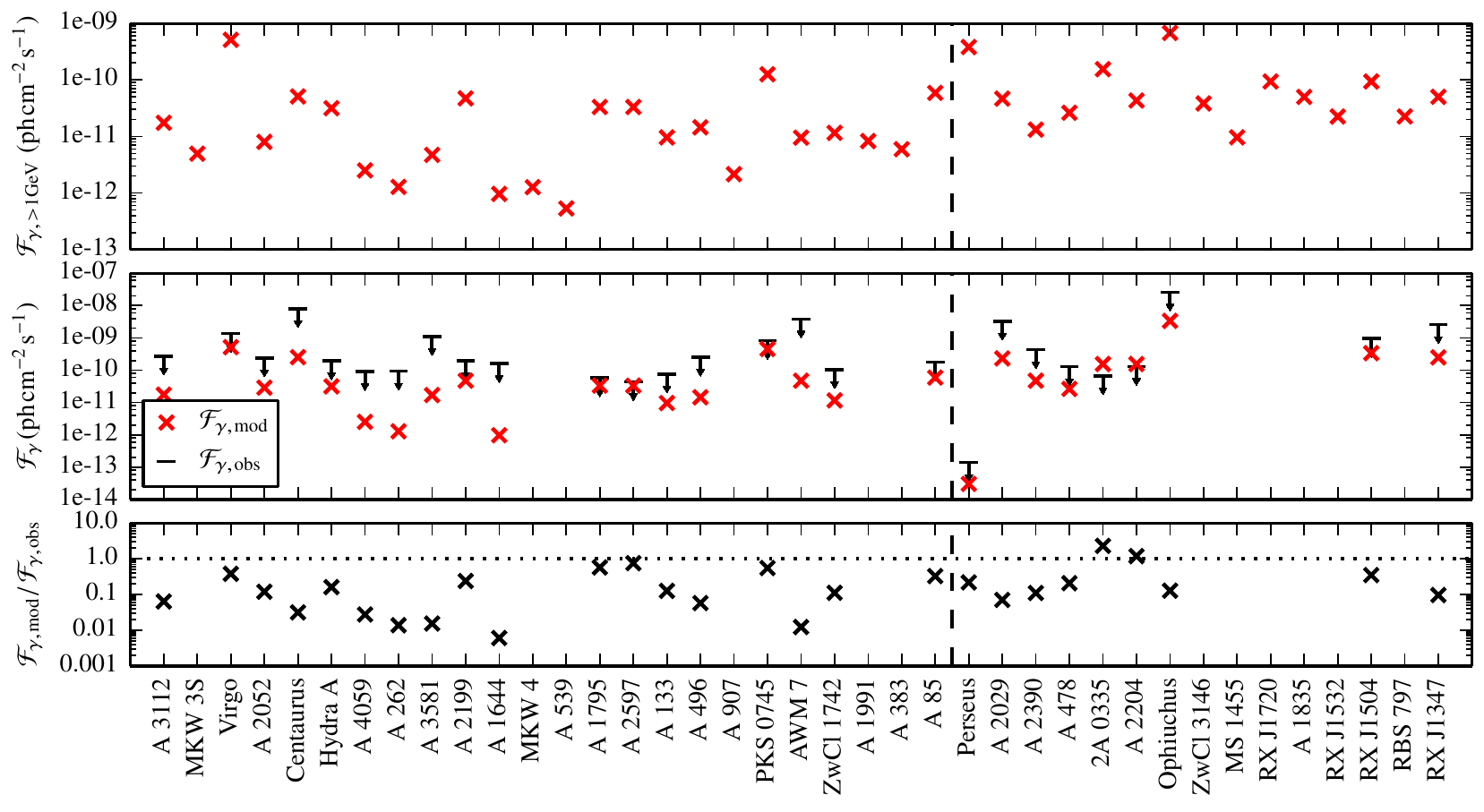}
  \caption{Comparison between the predicted gamma-ray flux as a result of
    hadronically induced pion decay and the constraints from observations. The
    top panel shows the predicted flux above $1\;\rmn{GeV}$ for all
    clusters. \C{In the middle panel we compare our predictions to the upper
      limits from \citet{Fermi2014}, \citet{Fermi2010},
      \citet[][Perseus]{Magic2012Perseus} and \citet{Dutson2013} and the
      gamma-ray detection in the Virgo cluster \citep{Abdo2009M87}.} We always
    compute the gamma-ray flux in the data-equivalent energy band (see
    Table~\ref{tab:sample} for details). The bottom panel illustrates the ratio
    between the predicted flux and the upper limits, indicating that present-day
    gamma-ray observations are not sensitive enough to seriously challenge the
    CR heating model (with the exceptions of 2A 0335 and A~2204, in which the CR
    heating model can be excluded based on gamma-ray observations).}
    \label{fig:gammaplot}
\end{figure*}

Hadronic interactions between CRs and thermal protons produce neutral pions
that decay into gamma-ray photons with a distinctive spectral signature in the
differential source function that peaks at energies of half the pions' rest
mass. We use upper limits to the extended gamma-ray emission of galaxy clusters
from \textit{Fermi} and MAGIC to probe our model.

%----------------------------------------------------------------------------------------------------
\subsection{Pion decay luminosity}

We follow \citet{Pfrommer2008} to determine the gamma-ray fluxes. Here,
$s_\gamma(E_\gamma)$ denotes the gamma-ray source function as a function of
energy. The omnidirectional integrated gamma-ray source density between two
energies $E_1$ and $E_2$ in units of photons per energy, per unit time, and per
unit volume is then given by
\begin{equation}
\lambda_{\gamma} = \int_{E_1}^{E_2} \rmn{d} E_{\gamma} s_{\gamma}(E_{\gamma}).\label{equ:gammasource}
\end{equation}
A detailed description of this formalism and the source function can be found in
Appendix~\ref{sec:appgamma}.  Integrating $\lambda_{\gamma}$ over the cluster
volume yields the photon luminosity per unit time
\begin{equation}
  \mathcal{L}_\gamma = \int \rmn{d} V \lambda_\gamma
  = 4 \upi \int_0^{r_\rmn{max,\gamma}} \lambda_\gamma r^2\rmn{d}r,
  \label{eq:gammal}
\end{equation}
where we use $r_\rmn{max,\gamma}=\max\left\{r_\rmn{out},200\rmn{~kpc}\right\}$
as the upper integration limit.  While the gamma-ray luminosity scales with
cluster mass, there is an enormous range in non-thermal luminosity at fixed mass
due to the large variance in gas density across our sample (see
Fig.~\ref{fig:LnonthermalvsMass}).  The latter effect dominates the variance of
the gamma-ray luminosity in our core sample as we quantitatively discuss in
Appendix~\ref{sec:applum}. We obtain the gamma-ray fluence from the luminosity
via
\begin{equation}
  \mathcal{F}_\gamma
  = \frac{\mathcal{L}_\gamma}{4 \upi D_\rmn{lum}^2}
  \label{eq:gammaf}
\end{equation}
with the luminosity distance $D_\rmn{lum}$.

\subsection{Comparison with gamma-ray limits}

We show the gamma-ray fluence above $1\rmn{~GeV}$ for all clusters in the upper
panel of Fig.~\ref{fig:gammaplot} (see Table~\ref{tab:sample} for numerical
values). The values are spread over three orders of magnitude between $10^{-12}$
and $10^{-9}\rmn{~ph\,cm^{-2}\,s^{-1}}$.  The fluences of clusters with an RMH
are somewhat higher that for cluster without an RMH, with median values of
$4\times10^{-11}$ and $1\times10^{-11}\rmn{~ph\,cm^{-2}\,s^{-1}}$, respectively.
This difference is smaller than the difference in gamma-ray luminosity of the
two subsamples, because RMH clusters are on average at higher redshifts that
partially compensates the larger luminosity (see
Fig.~\ref{fig:LnonthermalvsMass}).

Additionally, we compare our model fluences to observations. To this end, we
employ the upper limits from \citet{Fermi2010} who analyse data from the
\textit{Fermi} satellite for individual clusters. We also consider stacked
\textit{Fermi} limits provided by \citet{Fermi2014}\footnote{Note that the
  stacked \textit{Fermi} limits on individual cluster by \citet{Fermi2014}
  assume universality of the CR distribution as a result of diffusive shock
  acceleration at cosmological formation shocks \citep{Pinzke2010}. If the
  dominant CR population in clusters is injected by AGNs rather than by
  structure formation shocks, the limits may be somewhat weaker.} and
\citet{Dutson2013}, and we use \textit{Fermi} observations of the Virgo cluster
\citep{Abdo2009M87} as well as MAGIC observations of the Perseus cluster
\citep{Magic2012Perseus}. \C{Note that all values are upper limits except for the
Virgo cluster/M87.}

Since these authors report their upper limits for different energy bands, we
have to choose a data-equivalent energy band from $E_1$ to $E_2$ in
Equation~\eqref{equ:gammasource}. In the middle panel of
Fig.~\ref{fig:gammaplot} we compare those observational gamma-ray limits to our
predictions and show the ratio of predicted-to-expected gamma-ray emission in
the bottom panel (upper limits and data-equivalent energy ranges are shown in
Table~\ref{tab:sample}). While the expectations for most clusters are below the
upper limits, there are two clusters (2A 0335 and A~2204) that exceed the
observational constraints. In those clusters, we can exclude the CR heating
model based on gamma-ray observations alone. However, both of these clusters
host an RMH for which our model is already excluded by the radio data. Hence, we
conclude that while gamma-ray predictions come close to observational limits,
present-day gamma-ray observations are not sensitive enough to seriously
challenge the CR heating model.

Notable are the results for the Virgo cluster. \citet{Pfrommer2013} constructs a
CR population that simultaneously matches the observed gamma-ray emission and is
able to stably balance radiative cooling while adopting a {\em constant}
CR-to-thermal pressure ratio $X_\CR$ throughout the observed radio micro-halo
(i.e., for $r<35$~kpc). Our steady-state model has also been constructed to
offset radiative cooling but falls short of the observed gamma-ray emission by a
factor of $2.6$. This is mainly because conductive heating starts to balance
radiative cooling in our steady-state solution at radii $r\gtrsim20$~kpc and
causes the $X_\CR$ profile to steeply drop at this radius. Hence, the resulting
hadronic gamma-ray emission falls short of the value it would have if conductive
heating were absent. Moreover, in this work, we employ a slightly higher magnetic
field, which translates to a slightly lower CR pressure for the identical
heating rate, and a different cooling profile (which we infer from the ACCEPT
data base).

The second cluster that has been studied in detail is the Perseus cluster. Here,
we compare our model to TeV gamma-ray observations. At these energies the flux
from the central galaxy NGC~1275 has dropped significantly so that gamma-rays
from decaying pions should become dominant \citep{Magic2012Perseus}.  The chosen
energy range also explains the small absolute values for the gamma-ray fluence
in Perseus. Although our model agrees with the current limits, we note that
possible spectral steepening associated with CR streaming \citep{Wiener2013}
could weaken the MAGIC gamma-ray limit that assumes a single power-law spectrum
to TeV energies \citep{MAGIC2016}.

%%%%%%%%%%%%%%%%%%%%%%%%%%%%%%%%%%%%
\section{Emerging picture}
\label{sec:picture}

\subsection{A self-regulated scenario for CR heating, cooling, and star formation}

What is the conclusion of this at first sight disparate result that CR heating
is excluded as the predominant source of heating in clusters that manifestly
show non-thermal emission in form of RMHs? Let us summarize the main findings:

\begin{enumerate}
  \item Our steady-state solutions of \citetalias{Jacob2016a} demonstrate that
    radiative cooling can be balanced by CR heating in the central region and by
    thermal conduction in the outer region. The resulting CR-to-thermal pressure
    in the central region attains values of $X_\CR\approx0.05-0.1$ for clusters
    without an RMH, and shows systematically higher values of
    $X_\CR\approx0.1-0.25$ for clusters with RMHs.
  \item The level of hadronic radio and gamma-ray fluxes of our steady-state
    solutions is higher in clusters hosting an RMH because of the higher target
    density in RMH clusters (see Fig.~\ref{fig:slopes}) and excluded by observed
    NVSS and RMH fluxes.
  \item In contrast, the predicted non-thermal emission
    is below observational radio and gamma-ray data in cooling galaxy clusters
    without RMHs (with the exception of A 383 and A 85).
  \item Most importantly, the ratio of predicted-to-observed NVSS flux is
    dramatically increased in RMH clusters, the median of the flux ratio for
    both populations differs by a factor of a few hundred. In addition to the
    increased secondary flux noted in point (ii), the radio emission of the
    central AGN in clusters {\em without} a detected RMH is on average also much
    stronger. Because the AGN radio emission is a proxy for CR injection, this
    implies a significantly increased CR yield in the centre of those
    clusters. In particular, the predicted-to-observed NVSS flux ratio shows a
    continuous sequence from $10^{-4}$ at the lower end of non-RMH clusters to
    100 for the upper end of RMH clusters (bottom panel in
    Fig.~\ref{fig:nvssplot}).
\end{enumerate}

These different findings can be put together in form of a {\em self-regulation
  scenario of AGN feedback} in CC clusters for which we will provide
further evidence below. A strong AGN radio emission signals the abundant
injection of CRs into the centre.\footnote{Equipartition arguments for
  radio-emitting lobes demonstrate that the sum of CR electrons and magnetic
  fields can only account for a pressure fraction of $\simeq10\%$ in comparison
  to the ambient ICM pressure, with which the lobes are in approximate hydrostatic
  equilibrium \citep{Blanton2003,deGasperin2012}. This makes a
  plausible case for CR protons to supply the majority of internal energy of the
  bubbles \citep[see also][]{Pfrommer2013}.}  As these CRs stream
outwards they can balance radiative cooling via Alfv{\'e}n wave heating in the
central regions while conductive heating takes over at larger radii. Here, the
streaming CRs can heat the ICM homogeneously and locally stable
\citep{Pfrommer2013} by generating resonantly Alfv{\'e}n waves so that mass
deposition rates drop below $1~\rmn{M}_\odot~\rmn{yr}^{-1}$
\citepalias{Jacob2016a}.

Observationally, these CR heated systems could be associated with CC clusters
that do not have an observable {\em radio mini halo}. Instead, we predict a new
class of {\em radio micro halos}, that is associated with the radio synchrotron
emission of primary and secondary CR electrons surrounding the central AGN. {\em
  Radio micro halos} have thus far escaped detection due to the small extent of
the {\em micro halo} up to a few tens of kpcs and the large dynamic flux range
of the jet and halo emission. An exception that supports this hypothesis is the
only known {\em micro halo} in M87, the centre of the Virgo cluster, which can
only be observed due to its close proximity of 17 Mpc. The expected hadronic
gamma-ray emission can be identified with the low state of M87 \citep[][see also
  Fig.~\ref{fig:gammaplot}]{Pfrommer2013}.

Once the CR population has streamed sufficiently far from the centre and lost
enough energy in exciting Alfv{\'e}n waves, the gas cooling rate increases to
values above $1~\rmn{M}_\odot~\rmn{yr}^{-1}$ that should also fuel star
formation.  Hence, this picture would predict enhanced levels of star formation
in clusters in which CR heating ceases to be efficient, namely in those that are
hosting an RMH.  Our self-regulation scenario of CR-induced heating not only
predicts stably heated clusters on the one side and cooling systems with
abundant star formation on the other side, but also systems transitioning from
one state to the other, such as the Perseus cluster, A~85, or A~383.

\subsection{Supporting evidence for this picture}

\begin{figure*}
  \includegraphics{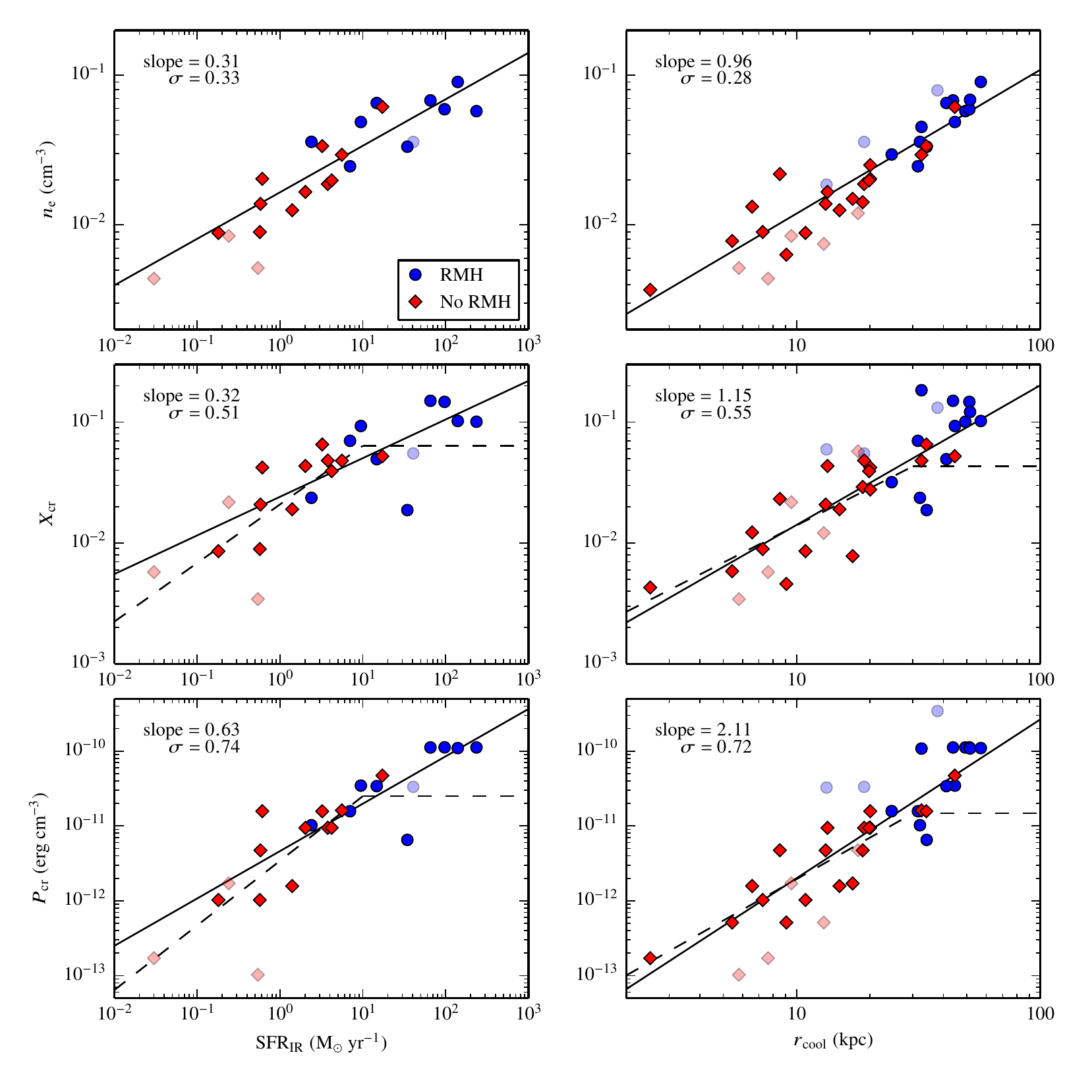}
  \caption{Relations between density (top), ratio of CR-to-thermal pressure
    (middle) and CR pressure (bottom) at a reference radius of $30\rmn{~kpc}$
    with the observed IR SFRs (left) and the cooling radius (right). The black
    lines are best-fitting power-law relations to our cluster core sample (full
    coloured data points). The more transparent data points denote clusters at
    the low- and high-mass end of our sample and are only shown for visual
    purposes.  The slope and the vertical log-normal scatter $\sigma$ of the fit
    are indicated in the upper left of each panel. RMH clusters populate
    the upper end of these correlations, which is characterized by large SFRs
    and cooling radii. The top two panels represent purely observational
    correlations while the middle and bottom panels employ the CR and thermal
    pressure profiles of the steady-state solutions of
    \citetalias{Jacob2016a}. Since for some clusters those solutions are
    excluded by radio data (see text), the dashed lines show fits to the
    remaining data points.}
    \label{fig:slopes}
\end{figure*}

\begin{table}
  \caption{Fit results for the correlations of the unbiased sample shown in
    Fig.~\ref{fig:slopes} using a power-law relation of the form $y(x) = a x^b$,
    where $y(x)$ is specified in the first column.$^{(1)}$}
\label{tab:slopefits}
\begin{tabular}{r  r r r r r r}
\hline
&\multicolumn{1}{c}{$a$} & \multicolumn{1}{c}{$b$} & \multicolumn{1}{c}{$\sigma$} \\
\hline
$ n_\rmn{e}$(SFR)\hphantom{$_\rmn{val}$}  	&$(1.6 \pm 0.2) \times 10^{-2\hphantom{0}}$	&$0.31 \pm 0.04$	&$0.33$\\
$ kT$(SFR)\hphantom{$_\rmn{val}$} 			&$(3.6 \pm 0.2) \hphantom{\; \times \,10^{-13}}$	&$0.06 \pm 0.02$	&$0.18$\\
$X_\rmn{cr}$(SFR)\hphantom{$_\rmn{val}$}	&$(2.4 \pm 0.4) \times 10^{-2\hphantom{0}}$	&$0.32 \pm 0.06$	&$0.51$\\
$X_\rmn{cr}$(SFR)$_\rmn{val}$				&$(2.1 \pm 0.3) \times 10^{-2\hphantom{0}}$	&$\hphantom{0}0.5 \pm 0.1\hphantom{0}$&$0.37$\\
$P_\rmn{cr}$(SFR)\hphantom{$_\rmn{val}$}	&$\hphantom{.0}(5 \pm 1) \hphantom{.0} \times 10^{-12}$	&$0.63 \pm 0.09$	&$0.74$\\
$P_\rmn{cr}$(SFR)$_\rmn{val}$				&$(3.4 \pm 0.8) \times 10^{-12}$					&$\hphantom{0}0.9 \pm 0.2\hphantom{0}$	&$0.57$\\
\hline
$ n_\rmn{e}$ ($r_\rmn{cool}$)\hphantom{$_\rmn{val}$} 	&$(1.3 \pm 0.3) \times 10^{-3\hphantom{0}}$	&$0.96 \pm 0.07$	&$0.28$\\
$ kT$ ($r_\rmn{cool}$)\hphantom{$_\rmn{val}$} 			&$(2.5 \pm 0.4) \hphantom{\;\times\, 10^{-4\hphantom{0}}} $ &$ 0.15 \pm 0.05$ &$ 0.19$\\
$X_\rmn{cr}$($r_\rmn{cool}$)\hphantom{$_\rmn{val}$}	&$(1.0 \pm 0.4) \times 10^{-3\hphantom{0}}$	&$\hphantom{0}1.2 \pm 0.1\hphantom{0}$ 	&$0.55$\\
$X_\rmn{cr}$($r_\rmn{cool}$)$_\rmn{val}$	&$(1.3 \pm 0.7) \times 10^{-3\hphantom{0}}$	&$\hphantom{0}1.0 \pm 0.2\hphantom{0}$ 	&$0.52$\\
$P_\rmn{cr}$($r_\rmn{cool}$)\hphantom{$_\rmn{val}$}	&$(1.5 \pm 0.8) \times 10^{-14}$					&$\hphantom{0}2.1 \pm 0.2\hphantom{0}$	&$0.72$\\
$P_\rmn{cr}$($r_\rmn{cool}$)$_\rmn{val}$	&$\hphantom{0}(3 \pm 2)\hphantom{.0} \times 10^{-14}$					&$\hphantom{0}1.8 \pm 0.3\hphantom{0}$	&$0.68$\\
\hline
\end{tabular}\\
(1) These fits were performed in logarithmic space, the scatter $\sigma$ was
obtained assuming a normal distribution for the deviation of the logarithm of
the data to the mean relation. SFRs are given in $\rmn{M_\odot\;yr^{-1}}$ and
cooling radii in kpc. Densities are measured in $\rmn{cm^{-3}}$, temperatures in
keV and CR pressures in $\rmn{erg\;cm^{-3}}$. The subscript ``val'' indicates
the relations of the subsample of clusters for which our model is valid (dashed
lines in Fig.~\ref{fig:slopes}).
\end{table}

To test this hypothesis, we scrutinize the cluster profiles for signs of such a
cycle. To this end, we study observed quantities such as densities and SFRs as
well as quantities that are predicted by the steady-state solutions such as the
required CR pressures to balance radiative cooling.

First, we correlate the observed electron number density at a reference radius
of $30\rmn{~kpc}$ to the observed SFRs (top left panel of Fig.~\ref{fig:slopes})
and the cooling radius (top right panel). We find clear correlations of the form
$n_\rmn{e}\propto \rmn{SFR}^{0.31}$ and $n_\rmn{e}\propto
r_{\rmn{cool}}^{0.96}$. The log-normal scatter of these relations is
$\sigma=0.33$ and $0.28$, respectively (see Table~\ref{tab:slopefits}, for the
fit parameters of the relation). Note that we exclude clusters at the low- and
high-mass end of our sample (shown with transparent colours) for the fit.  Most
importantly, clusters hosting an RMH populate the upper end of the correlation
that is characterized by the largest SFRs and cooling radii, i.e., {\em RMHs signal
cluster cores with enhanced cooling activity.}

In order to connect these empirical findings to our theoretically motivated
steady state solutions, we also determine the ratio of CR-to-thermal pressure
inferred from our steady state solutions \citepalias{Jacob2016a} at a reference
radius of $30\rmn{~kpc}$ and correlate it to the observed SFRs and the cooling
radius (middle panels of Fig.~\ref{fig:slopes}). We see a correlation that has a
similar dependence on SFR and $r_\rmn{cool}$, albeit with a larger scatter.
Dashed lines indicate the relations if only clusters are considered in which our
model is valid, i.e., if we exclude clusters that host an RMH as well as A~383
and A~85.  With the smaller sample, the relation is somewhat steeper for the
SFRs but remarkably similar for the cooling radius. Clusters with an RMH require
higher values of $X_\rmn{cr}$ than clusters without RMHs to balance the enhanced
cooling rates.

Last, we relate the CR pressure from the steady-state solutions at a reference
radius of $30\rmn{~kpc}$ to the observed SFR and cooling radius (bottom panels
of Fig.~\ref{fig:slopes}).  Since $P_\rmn{cr} \propto X_\rmn{cr} n_\rmn{e} kT$ and the
correlation of $kT$ with SFR and cooling radius shows no clear trends (see
Table~\ref{tab:slopefits}), we expect that the dependence of the CR pressure on
SFR and cooling radius derive from the previous relations. Indeed, we obtain
such steeper relations with a slope that is approximately given by the sum of
the slopes for the density and $X_\rmn{cr}$ relations. We find values of $0.63$
and $2.11$ for the scaling of $P_\rmn{cr}$ with SFR and cooling radius,
respectively.  However, the correlations of the CR pressures show the largest
scatter. As expected, clusters with an RMH have higher values for the CR
pressure than clusters without RMHs. Dashed lines indicate again the results for
the sample in which our steady-state solutions are in agreement with the
observational radio (and gamma-ray) data.

\begin{figure}
  \includegraphics{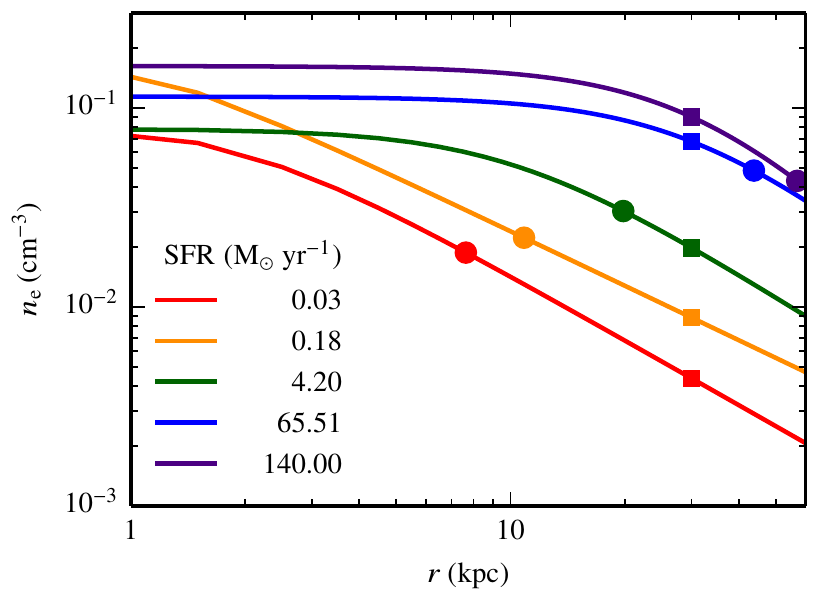}
  \caption{Representative electron number density profiles of five clusters with
    different SFRs, which are distributed along the correlation shown in
    Fig.~\ref{fig:slopes}. Note that the cluster with the lowest SFR is at the
    low-mass end and not part of our cluster core sample. The squares indicate
    the density at a reference radius of $30\rmn{~kpc}$ whereas the circles
    denote the density at the cooling radius, $r_{\rm cool}$, of these systems.}
    \label{fig:density4clusters}
\end{figure}

In order to interpret these relations, we show in
Fig.~\ref{fig:density4clusters} the fit to the density profiles of five
representative clusters along the correlations shown in Fig.~\ref{fig:slopes},
with a wide distribution in SFRs (RX J1504.1, ZwCl~3146, A~3112, Centaurus,
MKW~4, moving from high to low SFRs). Note that the cluster with the lowest SFR
is not part of our core sample due to its low virial mass. For each cluster the
squares indicate the density at the reference radius of $30\rmn{~kpc}$ and the
circle marks the cooling radius $r_\rmn{cool}$.  Clearly, higher densities imply
larger cooling rates and thus larger cooling radii. This puts a higher demand on
the heating rate to balance the much increased cooling rate. Because these
higher densities correlate with an increased SFR, the balance is apparently
unsuccessful.  This implies that these clusters are currently not stably heated
but can cool to some extent.  Hence, it might not be necessary for potential
heating mechanisms to (fully) counteract radiative cooling in those clusters.

\begin{figure*}
  \includegraphics{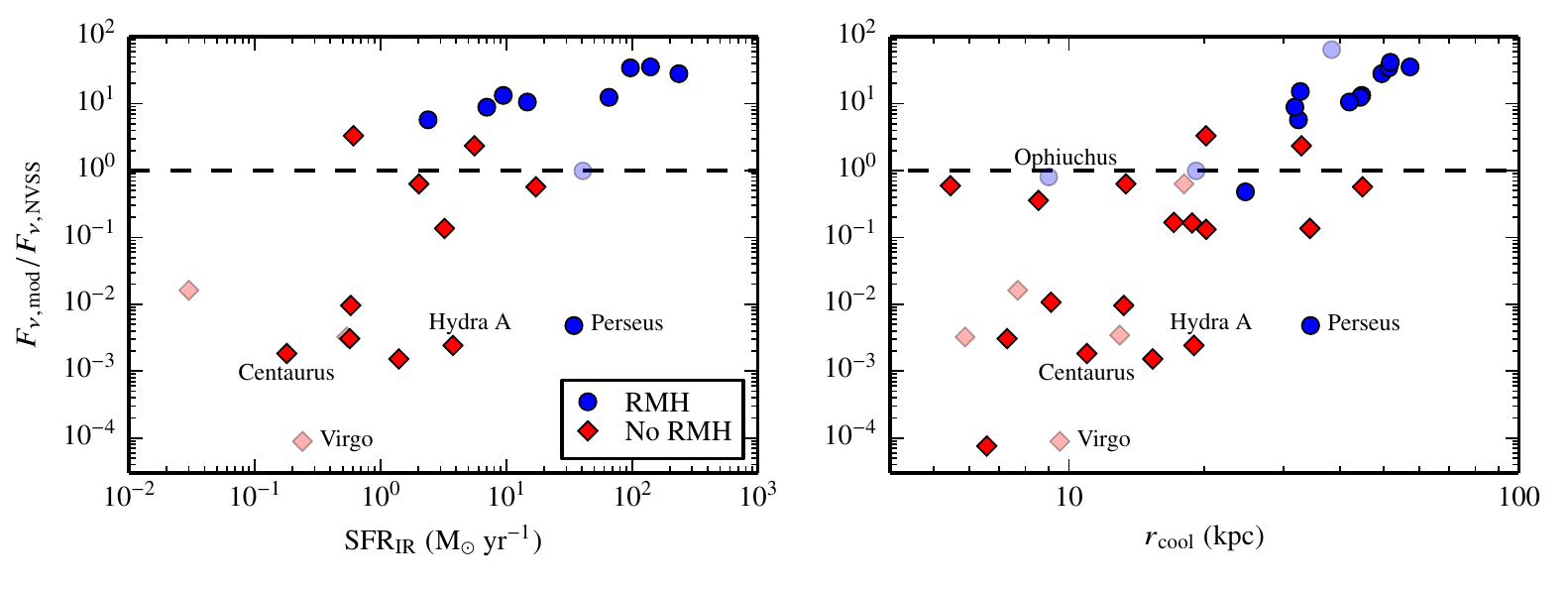}
  \caption{A measurement of the applicability of our model is the ratio of the
    modelled-to-observed NVSS flux. The modelled 1.4~GHz radio flux derives from
    the hadronically generated synchrotron emission of our steady-state CR
    population that stably balances radiative cooling.  Here, we compare the
    flux ratio to the observed SFRs (left) and cooling radii (right) and
    separate clusters with and without an RMH by colour. Clusters with higher
    SFRs can not be successfully heated by streaming CRs while this is a likely
    possibility for clusters with lower SFRs, as expected for a self-regulated
    heating-cooling cycle.}
    \label{fig:population}
\end{figure*}

As we demonstrate, CR heating is a prime candidate for providing the necessary
heating rate: clusters with low SFRs can be CR heated unlike clusters with high
SFRs. This is emphasized in Fig.~\ref{fig:population} where we compare the ratio
of modelled radio flux-to-NVSS flux with the SFR (left) and with the cooling
radius (right).  The figure shows that the flux ratio increases with SFR and
cooling radius. Since the ratio of predicted-to-observed radio flux is a measure
for the applicability of our model, this demonstrates that CR heating is viable
in clusters with low SFRs and not applicable in clusters with higher SFRs. These
results support the picture of a CR heating--radiative cooling cycle.

A note on time-scales is in order since our picture requires that the density
profile of the clusters is transformed within a heating cycle.  The density
profile can only rearrange itself on a dynamical (free-fall) time,
$\tau_\rmn{ff} = \sqrt{3 \upi/(32 G \rho)} \approx 7 \times 10^{7}\rmn{~yr}$,
assuming a typical total mass density of $\rho =9 \times
10^{-25}\rmn{~g\,cm^{-3}}$. (We obtained this density scale by solving the
equation for hydrostatic equilibrium of our steady-state solutions.)  This time-scale is of the same order as typical AGN duty cycles, which range from a few
times $10^7~\rmn{yr}$ to a few times $10^8~\rmn{yr}$ \citep{1987MNRAS.225....1A,2005Natur.433...45M,
  2005ApJ...628..629N, 2008MNRAS.388..625S}.  One could
imagine that the rearrangement of the density profile is modulated by a few to
several short-duration AGN feedback cycles that maintain a quasi-steady CR flux
on the longer time-scale. We will study the consequences of these considerations
in future work using numerical three-dimensional magneto-hydrodynamical
simulations with CR physics that is coupled to AGN feedback
\citep{Pfrommer2016}.

\C{ Despite these favourable results for a CR regulated feedback cycle, we can
  not exclude that such a cycle can be driven by another heating mechanism like
  mixing \citep{Brueggen2002, Hillel2016, Yang2016b}, sound or shock waves
  \citep{Fabian2003, Fabian2006, Fabian2017} although similarly thorough
  statistical studies as we present here would have to be conducted for the
  alternative scenarios.}

\subsection{Origin of RMHs}

We saw that RMHs are lighthouses signalling an increased cooling and SFR in CC clusters. Is there also a causal connection between
RMHs and increased cooling rates? While we have seen that streaming CRs are not
abundant enough in the radio emitting volume of RMHs to balance radiative cooling,
they could still be energetic enough to power the observed radio emission via
the injection of secondary electrons.

To test this hypothesis, we take the spatial CR pressure profile of our steady
state solution of a non-RMH cluster that is just compatible with being CR heated
and on its way to become a transitional object. Such clusters are characterized
by a comparably large CR-to-thermal pressure ratio of $X_\CR \approx 0.06$
(Fig.~\ref{fig:slopes}). As the cluster is transforming into a stronger cooling
CC system, the CR population is transported outwards by streaming.

Additionally, a large number of CC clusters show spiral contact discontinuities
in the X-ray surface brightness maps, indicating sloshing or swirling gas
motions induced by minor mergers, and implying also advective CR transport by
turbulence \citep{Markevitch2007, Simionescu2012, ZuHone2013}.  Advective
compression or expansion by means of gas motions yield adiabatic gains or losses
of the CR distribution, respectively. Interestingly, the process of CR streaming
is also a purely adiabatic process from the perspective of the CRs
\citep{Ensslin2011, Pfrommer2016}. While dissipation of the excited Alfv{\'e}n
waves is not a reversible process, the energy transferred to the wave fields
originates from adiabatic work done by the expanding CR population on the wave
frame.

To estimate the net CR pressure losses during the outwards streaming and
formation of RMHs, we only need to consider the adiabatic CR losses across a
density contrast $\delta$, which is given by
\begin{equation}
  \label{eq:adiabatic}
  P_{\CR,2} = P_{\CR,1} \delta^{\gamma_\CR}.
\end{equation}
This implies a decrease of the CR pressure (in the Lagrangian wave frame) by a
factor ranging from 2.5 to 20 for a density contrast $\delta=0.5$ -- 0.1.  We
cannot uniquely relate this result to the change of CR pressure at a fixed point
in space, since this depends on the time-dependent injection rate of CRs by the
AGN at the centre and on the ratio of streaming-to-turbulent advection time-scales $\gamma_{\rmn{tu}}=\tau_{\rmn{st}}/\tau_{\rmn{tu}}$
\citep{Ensslin2011}. Without further driving the sloshing motions that drive
turbulent advection start to cease and streaming becomes more important in
comparison to advection such that $\gamma_{\rmn{tu}}$ drops. If we assume
negligible central injection, the outwards streaming CRs cause the CR pressure
profile to flatten. However, the steep density profiles of CCs translate into
steep CR pressure profiles, which remain steep despite the increasing importance
of streaming. Even a value of $\gamma_{\rmn{tu}}=2$ shows an almost invariant CR
profile (see fig.~1 in \citealt{Zandanel2014}) and thus, the shape of the
$X_\CR$ profile remains approximately constant.  This might explain how the
approximately constant $X_\CR$ profiles of our steady state solutions can be
transformed into the equally flat $X_\CR$ profiles that are inferred from the
emission profiles of RMHs \citep{Pfrommer2004c, Zandanel2014}.  As a result, the
CR-to-thermal pressure ratio $X_\CR$ at a given point in space is expected to
drop by a factor of a few to about 100, depending on the time-dependent CR
injection rate and $\gamma_{\rmn{tu}}$. This range is in line with estimates for
RMHs, that require values of $X_\CR=3\times10^{-4}$ (Ophiuchus) to 0.02
(Perseus, see figure 2 in \citealt{Zandanel2014}). This plausibility estimate
suggests that RMHs could be powered hadronically by CRs that have heated the
cluster core in the past.

We complement these energetic estimates of CR streaming by calculating spectra
of RMHs and our predicted radio micro halos. Similar to the flux calculations in
Section~\ref{sec:radio}, we first project the emissivity along the line of
sight, assuming a radial extent of $r_\rmn{max,\parallel} = \max
\left\{r_\rmn{out}, 200\rmn{~kpc} \right\}$. In contrast to the previous
calculations, here we cut out a hollow cylinder with inner radius $r_\rmn{min,
  \bot} =2.5 \rmn{~kpc}$ and outer radius $r_\rmn{max, \bot} =
\min\left\{r_\rmn{RMH}, r_\rmn{max, \parallel}\right\}$. Note that here we adopt
$r_\rmn{RMH} = 34\rmn{~kpc}$ for the Virgo cluster \citep{deGasperin2012}. This
procedure attempts to mock observational determinations of RMH fluxes, which are
often dominated in the cluster centre by the radio jet emission.  The outer
radius is chosen such that it mimics the extent of observed RMHs.

\begin{figure*}
  \includegraphics{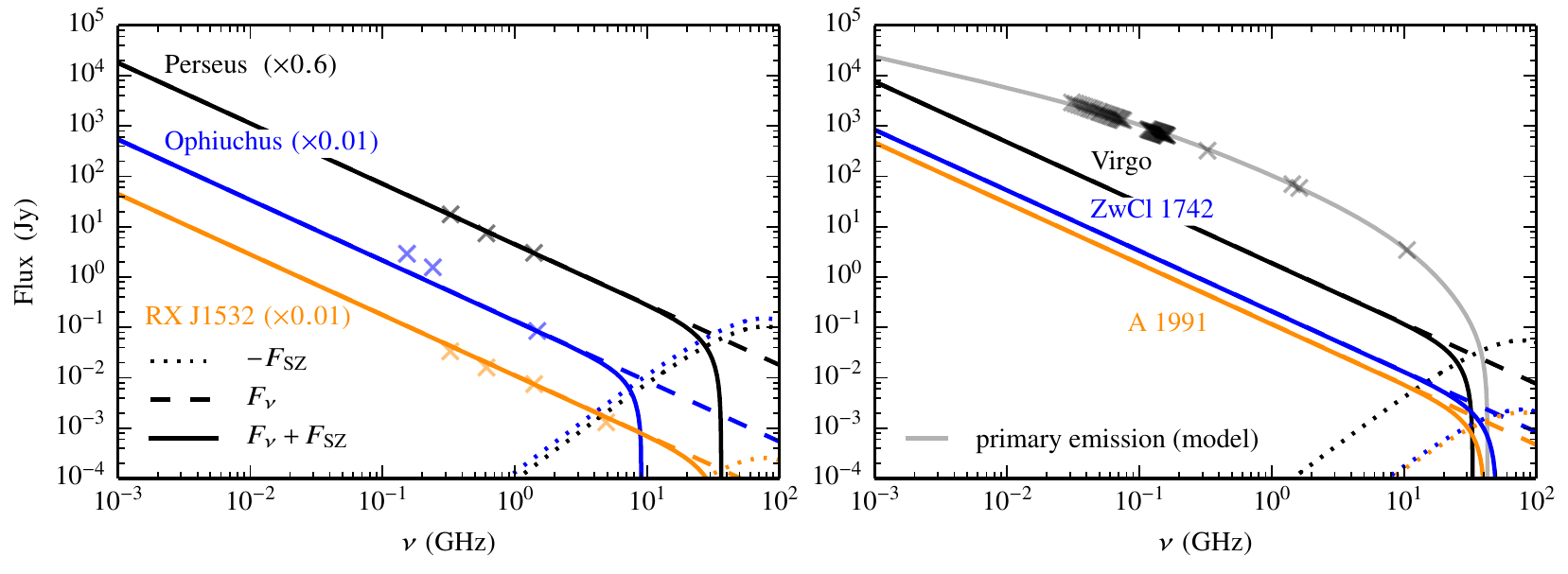}
  \caption{Spectra of three RMHs (left) and three predicted radio micro halos
    (right).  The data for the RMHs are taken from
    \citet[][Perseus]{Sijbring1993}, \citet[][Ophiuchus]{Murgia2010} and
    \citet[][RX J1532]{Giacintucci2014}, and the data for M87 is taken from the
    halo region of \citet{deGasperin2012}; the grey spectrum is the modelled
    primary synchrotron emission, assuming a continuous injection that was
    switched off after a certain time \citep{deGasperin2012,
      Pfrommer2013}. Dashed lines correspond to unattenuated RMH fluxes, scaled
    to the 1.4 GHz flux by a scaling factor indicated in the left-hand panel. Dotted
    lines show the negative flux decrement due to the thermal Sunyaev--Zel'dovich
    effect. This induces a cut-off to the observable radio spectrum, indicated by
    the solid lines.}
    \label{fig:spectrum}
\end{figure*}

In Fig.~\ref{fig:spectrum}, we compare the resulting spectra of observed RMHs and
the predicted radio micro halos.  Dashed lines show the unattenuated radio
fluxes, scaled to the 1.4~GHz flux by a scaling factor indicated in the left-hand
panel. Dotted lines show the negative flux decrement due to the thermal
Sunyaev--Zel'dovich effect, which we determine as in
\citet{Ensslin2002}.\footnote{For assessing the impact of the Sunyaev--Zel'dovich
  effect on the radio spectra, we integrate the thermal electron pressure of the
  ICM over the same (hollow) cylinder as for the calculation of the emissivity,
  but attempt to extend it along the line of sight as far as possible. For
  practical reasons, this implies an integration limit of
  $r_\rmn{max,\parallel}$ for all clusters but Virgo because of its wide radio
  spectral coverage. In this cluster, we extend the electron population along
  the line of sight to $800\rmn{~kpc}$ and find that our density and temperature
  fits agree reasonably well with the ROSAT data at that radius
  \citep{Boehringer1994, Nulsen1995}.}  This induces a cut-off to the observable
RMH spectra, indicated by the solid lines. The radio micro halo of M87
\citep[black data points from][]{deGasperin2012} is presumably generated by
primary accelerated CR electrons that have escaped from the bubbles. This
component was modelled assuming a continuous injection that was switched off
after a certain time (grey solid line). This causes the spectrum to drop
exponentially above a break frequency, which corresponds to the cooling time
since the switch-off. Despite the harder intrinsic spectrum of the
hadronically induced secondary component (black solid line) in comparison to the
convex curvature of the primary component, the presence of the
Sunyaev--Zel'dovich cut-off precludes a detection of the subdominant hadronic
component in M87.

There is a significant range of radio mini and micro halo fluxes
(Figs~\ref{fig:rmhfluxesplot} and \ref{fig:spectrum}). Especially the
comparably tight range of RMH redshifts and thus luminosity distances also
implies a range in luminosities. This matches our picture in which RMHs serve as
sign posts of the upper end of a continuous sequence in cooling properties.  The
observed range of gas densities and CR pressures causes the observed diversity
of radio luminosities.

\section{Conclusions}
\label{sec:conclusions}

CR heating has recently re-emerged as an attractive scenario for mediating
energetic feedback by AGNs at the centres of galaxy clusters \citep{Guo2008,
  Ensslin2011, Fujita2011, Wiener2013, Pfrommer2013}. However, all theoretical
studies to date have concentrated on individual objects or a very small sample
size, precluding statistically sound conclusions on any heating model. 

In this sequence of two papers, we have selected a rich sample of 39 CC clusters
and \C{found steady-state solutions} of the hydrodynamic equations coupled to
the CR energy equation. In those, radiative cooling is balanced by thermal
conduction at large scales and by CR heating in the central regions.

We find that those solutions are ruled out in a subsample of clusters that host
RMHs because the predicted hadronically induced non-thermal emission exceeds
observational radio and (some) gamma-ray data. On the contrary, the predicted
non-thermal emission respects observational radio data in CC clusters without
RMHs (with the exception of A~383 and A~85, in which the CR-heating solution is
barely ruled out). Those non-RMH clusters show exceptionally large AGN radio
fluxes, which should be accompanied by an abundant injection of CRs and -- by
extension -- should give rise to a large CR heating rate.

This enables us for the first time to put forward a statistically rooted,
self-regulated model of AGN feedback. We propose that non-RMH clusters are
heated by streaming CRs homogeneously throughout the cooling region through the
generation and dissipation of Alfv{\'e}n waves. On the contrary, CR heating
appears to be insufficient to \C{fully} balance the enhanced cooling in RMH
clusters. These clusters are also characterized by large SFRs, questioning the
presence of a stable heating mechanism that balances the cooling rate.  In those
systems, thermal conduction should still regulate radiative cooling on large
scales, which however is unable to adjust to local thermal fluctuations in the
cooling rate because of the strong temperature dependence of the conductivity
and may give rise to local thermal instability. \C{However, there will still be
  some residual level of CR heating in those cooling systems that quenches
  radiative cooling but is not able to completely offset it}.

We emphasize that our self-regulation scenario of CR-induced heating not only
predicts stably heated clusters and cooling clusters with abundant star
formation, but also systems transitioning from one state to the other, a
prominent example of which appears to be the Perseus cluster.

We predict {\em radio micro halos} of scales up to a few kpcs surrounding the
AGNs of these CR-heated clusters, resembling the diffuse radio emission around
Virgo's central galaxy, M87. Once the CR population has streamed sufficiently
far from the centre, it has lost enough energy so that its heating rate is
unable to balance radiative cooling any more. As a result star formation
increases in clusters that we empirically identify to host an RMH. We suggest
that the CR population that has heated the cluster core in the past is now
injecting secondary electrons that power the RMH.

Our new picture makes a number of novel predictions that allow scrutinizing it.
\begin{enumerate}
  \item We predict the presence of {\em radio micro halos} associated with {\em
    all} CC clusters that host no classic RMH and have small SFRs \citep[or
    alternatively H$\alpha$ luminosities,][]{Voit2008}. While this secondary
    emission component is expected to have a harder spectrum in comparison to
    the convexly curved, primary radio emission, we find that the negative flux
    decrement owing to the thermal Sunyaev--Zel'dovich effect typically cuts
    these emission components off at high frequencies
    ($\nu\gtrsim10-50$~GHz). In Virgo, the primary emission component
    predominates the hadronically induced secondary emission at all observable
    radio emission frequencies. Hence, we envision the harder secondary emission
    to predominate the primary component only in those cases where the latter
    has already cooled sufficiently down, i.e., at late times after the release
    of the CR electrons from the bubbles or at larger cluster-centric radii.
  \item We predict an observable steady-state gamma-ray signal resulting from
    hadronic CR interactions with the ICM. The spectral index that is expected
    to be correlated to the injection (electron and proton) index that can be
    probed at small radii with low-frequency radio observations
    \citep{Pfrommer2013}.
\end{enumerate}

Future magneto-hydrodynamic, three-dimensional cosmological simulations that
follow CR physics are necessary to study possible time-dependent effects of the
suggested scenario such as the impact of CR duty cycles on the heating rates and
to address non-spherical geometries associated with the rising AGN bubbles.

\section*{Acknowledgements}

We thank Volker Springel for helpful suggestions, acknowledge enlightening
discussions with Torsten En{\ss}lin\C{, and an anonymous referee for a
  constructive report}.  SJ acknowledges funding through the graduate college
\textit{Astrophysics of cosmological probes of gravity} by
Landesgraduiertenakademie Baden-W\"urttemberg. CP acknowledges support by the
ERC-CoG grant CRAGSMAN-646955. Both authors have been supported by the Klaus
Tschira Foundation.

%%%%%%%%%%%%%%%%%%%%%%%%%%%%%%%%%%%%%%%%%%%%%%%%%%

%%%%%%%%%%%%%%%%%%%% REFERENCES %%%%%%%%%%%%%%%%%%

\bibliographystyle{mnras}
\bibliography{../bib}

\begin{thebibliography}{}
\makeatletter
\relax
\def\mn@urlcharsother{\let\do\@makeother \do\$\do\&\do\#\do\^\do\_\do\%\do\~}
\def\mn@doi{\begingroup\mn@urlcharsother \@ifnextchar [ {\mn@doi@}
  {\mn@doi@[]}}
\def\mn@doi@[#1]#2{\def\@tempa{#1}\ifx\@tempa\@empty \href
  {http://dx.doi.org/#2} {doi:#2}\else \href {http://dx.doi.org/#2} {#1}\fi
  \endgroup}
\def\mn@eprint#1#2{\mn@eprint@#1:#2::\@nil}
\def\mn@eprint@arXiv#1{\href {http://arxiv.org/abs/#1} {{\tt arXiv:#1}}}
\def\mn@eprint@dblp#1{\href {http://dblp.uni-trier.de/rec/bibtex/#1.xml}
  {dblp:#1}}
\def\mn@eprint@#1:#2:#3:#4\@nil{\def\@tempa {#1}\def\@tempb {#2}\def\@tempc
  {#3}\ifx \@tempc \@empty \let \@tempc \@tempb \let \@tempb \@tempa \fi \ifx
  \@tempb \@empty \def\@tempb {arXiv}\fi \@ifundefined
  {mn@eprint@\@tempb}{\@tempb:\@tempc}{\expandafter \expandafter \csname
  mn@eprint@\@tempb\endcsname \expandafter{\@tempc}}}

\bibitem[\protect\citeauthoryear{{Abdo} et~al.,}{{Abdo}
  et~al.}{2009}]{Abdo2009M87}
{Abdo} A.~A.,  et~al., 2009, \mn@doi [\apj] {10.1088/0004-637X/707/1/55}, \href
  {http://adsabs.harvard.edu/abs/2009ApJ...707...55A} {707, 55}

\bibitem[\protect\citeauthoryear{{Ackermann} et~al.,}{{Ackermann}
  et~al.}{2010}]{Fermi2010}
{Ackermann} M.,  et~al., 2010, \mn@doi [\apjl] {10.1088/2041-8205/717/1/L71},
  \href {http://adsabs.harvard.edu/abs/2010ApJ...717L..71A} {717, L71}

\bibitem[\protect\citeauthoryear{{Ackermann} et~al.,}{{Ackermann}
  et~al.}{2014}]{Fermi2014}
{Ackermann} M.,  et~al., 2014, \mn@doi [\apj] {10.1088/0004-637X/787/1/18},
  \href {http://adsabs.harvard.edu/abs/2014ApJ...787...18A} {787, 18}

\bibitem[\protect\citeauthoryear{{Ahnen} et~al.,}{{Ahnen}
  et~al.}{2016}]{MAGIC2016}
{Ahnen} M.~L.,  et~al., 2016, \mn@doi [\aap] {10.1051/0004-6361/201527846},
  \href {http://adsabs.harvard.edu/abs/2016A%26A...589A..33A} {589, A33}

\bibitem[\protect\citeauthoryear{{Aleksi{\'c}} et~al.,}{{Aleksi{\'c}}
  et~al.}{2012}]{Magic2012Perseus}
{Aleksi{\'c}} J.,  et~al., 2012, \mn@doi [\aap] {10.1051/0004-6361/201118502},
  \href {http://adsabs.harvard.edu/abs/2012A%26A...541A..99A} {541, A99}

\bibitem[\protect\citeauthoryear{{Alexander} \& {Leahy}}{{Alexander} \&
  {Leahy}}{1987}]{1987MNRAS.225....1A}
{Alexander} P.,  {Leahy} J.~P.,  1987, \mn@doi [\mnras]
  {10.1093/mnras/225.1.1}, \href
  {http://adsabs.harvard.edu/abs/1987MNRAS.225....1A} {225, 1}

\bibitem[\protect\citeauthoryear{{B{\^i}rzan}, {Rafferty}, {McNamara}, {Wise}
  \& {Nulsen}}{{B{\^i}rzan} et~al.}{2004}]{Birzan2004}
{B{\^i}rzan} L.,  {Rafferty} D.~A.,  {McNamara} B.~R.,  {Wise} M.~W.,
  {Nulsen} P.~E.~J.,  2004, \mn@doi [\apj] {10.1086/383519}, \href
  {http://adsabs.harvard.edu/abs/2004ApJ...607..800B} {607, 800}

\bibitem[\protect\citeauthoryear{{Blanton}, {Sarazin}  \& {McNamara}}{{Blanton}
  et~al.}{2003}]{Blanton2003}
{Blanton} E.~L.,  {Sarazin} C.~L.,   {McNamara} B.~R.,  2003, \apj, \href
  {http://esoads.eso.org/cgi-bin/nph-bib_query?bibcode=2003ApJ...585..227B&amp;db_key=AST}
  {585, 227}

\bibitem[\protect\citeauthoryear{{B{\"o}hringer}, {Briel}, {Schwarz}, {Voges},
  {Hartner}  \& {Tr{\"u}mper}}{{B{\"o}hringer} et~al.}{1994}]{Boehringer1994}
{B{\"o}hringer} H.,  {Briel} U.~G.,  {Schwarz} R.~A.,  {Voges} W.,  {Hartner}
  G.,   {Tr{\"u}mper} J.,  1994, \mn@doi [\nat] {10.1038/368828a0}, \href
  {http://adsabs.harvard.edu/abs/1994Natur.368..828B} {368, 828}

\bibitem[\protect\citeauthoryear{{Br{\"u}ggen} \& {Kaiser}}{{Br{\"u}ggen} \&
  {Kaiser}}{2002}]{Brueggen2002}
{Br{\"u}ggen} M.,  {Kaiser} C.~R.,  2002, \nat, \href
  {http://adsabs.harvard.edu/abs/2002Natur.418..301B} {418, 301}

\bibitem[\protect\citeauthoryear{{Cavagnolo}, {Donahue}, {Voit}  \&
  {Sun}}{{Cavagnolo} et~al.}{2009}]{Cavagnolo2009}
{Cavagnolo} K.~W.,  {Donahue} M.,  {Voit} G.~M.,   {Sun} M.,  2009, \mn@doi
  [\apjs] {10.1088/0067-0049/182/1/12}, \href
  {http://adsabs.harvard.edu/abs/2009ApJS..182...12C} {182, 12}

\bibitem[\protect\citeauthoryear{{Chandran} \& {Rasera}}{{Chandran} \&
  {Rasera}}{2007}]{2007ApJ...671.1413C}
{Chandran} B.~D.~G.,  {Rasera} Y.,  2007, \mn@doi [\apj] {10.1086/521619},
  \href {http://adsabs.harvard.edu/abs/2007ApJ...671.1413C} {671, 1413}

\bibitem[\protect\citeauthoryear{{Churazov}, {Br{\"u}ggen}, {Kaiser},
  {B{\"o}hringer}  \& {Forman}}{{Churazov} et~al.}{2001}]{2001ApJ...554..261C}
{Churazov} E.,  {Br{\"u}ggen} M.,  {Kaiser} C.~R.,  {B{\"o}hringer} H.,
  {Forman} W.,  2001, \mn@doi [\apj] {10.1086/321357}, \href
  {http://adsabs.harvard.edu/abs/2001ApJ...554..261C} {554, 261}

\bibitem[\protect\citeauthoryear{{Coble} et~al.,}{{Coble}
  et~al.}{2007}]{Coble2007}
{Coble} K.,  et~al., 2007, \mn@doi [\aj] {10.1086/519973}, \href
  {http://adsabs.harvard.edu/abs/2007AJ....134..897C} {134, 897}

\bibitem[\protect\citeauthoryear{{Colafrancesco} \&
  {Marchegiani}}{{Colafrancesco} \& {Marchegiani}}{2008}]{Colafrancesco2008}
{Colafrancesco} S.,  {Marchegiani} P.,  2008, \mn@doi [\aap]
  {10.1051/0004-6361:20078428}, \href
  {http://adsabs.harvard.edu/abs/2008A%26A...484...51C} {484, 51}

\bibitem[\protect\citeauthoryear{{Condon}, {Cotton}, {Greisen}, {Yin},
  {Perley}, {Taylor}  \& {Broderick}}{{Condon} et~al.}{1998}]{Condon1998}
{Condon} J.~J.,  {Cotton} W.~D.,  {Greisen} E.~W.,  {Yin} Q.~F.,  {Perley}
  R.~A.,  {Taylor} G.~B.,   {Broderick} J.~J.,  1998, \mn@doi [\aj]
  {10.1086/300337}, \href {http://adsabs.harvard.edu/abs/1998AJ....115.1693C}
  {115, 1693}

\bibitem[\protect\citeauthoryear{{Donahue}, {Sun}, {O'Dea}, {Voit}  \&
  {Cavagnolo}}{{Donahue} et~al.}{2007}]{Donahue2007}
{Donahue} M.,  {Sun} M.,  {O'Dea} C.~P.,  {Voit} G.~M.,   {Cavagnolo} K.~W.,
  2007, \mn@doi [\aj] {10.1086/518230}, \href
  {http://adsabs.harvard.edu/abs/2007AJ....134...14D} {134, 14}

\bibitem[\protect\citeauthoryear{{Dutson}, {White}, {Edge}, {Hinton}  \&
  {Hogan}}{{Dutson} et~al.}{2013}]{Dutson2013}
{Dutson} K.~L.,  {White} R.~J.,  {Edge} A.~C.,  {Hinton} J.~A.,   {Hogan}
  M.~T.,  2013, \mn@doi [\mnras] {10.1093/mnras/sts477}, \href
  {http://adsabs.harvard.edu/abs/2013MNRAS.429.2069D} {429, 2069}

\bibitem[\protect\citeauthoryear{{En{\ss}lin}}{{En{\ss}lin}}{2002}]{Ensslin2002}
{En{\ss}lin} T.~A.,  2002, \mn@doi [\aap] {10.1051/0004-6361:20021613}, \href
  {http://adsabs.harvard.edu/abs/2002A%26A...396L..17E} {396, L17}

\bibitem[\protect\citeauthoryear{{En{\ss}lin}, {Pfrommer}, {Springel}  \&
  {Jubelgas}}{{En{\ss}lin} et~al.}{2007}]{Ensslin2007}
{En{\ss}lin} T.~A.,  {Pfrommer} C.,  {Springel} V.,   {Jubelgas} M.,  2007,
  \mn@doi [\aap] {10.1051/0004-6361:20065294}, \href
  {http://adsabs.harvard.edu/abs/2007A%26A...473...41E} {473, 41}

\bibitem[\protect\citeauthoryear{{En{\ss}lin}, {Pfrommer}, {Miniati}  \&
  {Subramanian}}{{En{\ss}lin} et~al.}{2011}]{Ensslin2011}
{En{\ss}lin} T.,  {Pfrommer} C.,  {Miniati} F.,   {Subramanian} K.,  2011,
  \mn@doi [\aap] {10.1051/0004-6361/201015652}, \href
  {http://adsabs.harvard.edu/abs/2011A%26A...527A..99E} {527, A99}

\bibitem[\protect\citeauthoryear{{Ettori}, {Gastaldello}, {Leccardi},
  {Molendi}, {Rossetti}, {Buote}  \& {Meneghetti}}{{Ettori}
  et~al.}{2010}]{Ettori2010}
{Ettori} S.,  {Gastaldello} F.,  {Leccardi} A.,  {Molendi} S.,  {Rossetti} M.,
  {Buote} D.,   {Meneghetti} M.,  2010, \mn@doi [\aap]
  {10.1051/0004-6361/201015271}, \href
  {http://adsabs.harvard.edu/abs/2010A%26A...524A..68E} {524, A68}

\bibitem[\protect\citeauthoryear{{Fabian}, {Sanders}, {Allen}, {Crawford},
  {Iwasawa}, {Johnstone}, {Schmidt}  \& {Taylor}}{{Fabian}
  et~al.}{2003}]{Fabian2003}
{Fabian} A.~C.,  {Sanders} J.~S.,  {Allen} S.~W.,  {Crawford} C.~S.,  {Iwasawa}
  K.,  {Johnstone} R.~M.,  {Schmidt} R.~W.,   {Taylor} G.~B.,  2003, \mn@doi
  [\mnras] {10.1046/j.1365-8711.2003.06902.x}, \href
  {http://adsabs.harvard.edu/abs/2003MNRAS.344L..43F} {344, L43}

\bibitem[\protect\citeauthoryear{{Fabian}, {Sanders}, {Taylor}, {Allen},
  {Crawford}, {Johnstone}  \& {Iwasawa}}{{Fabian} et~al.}{2006}]{Fabian2006}
{Fabian} A.~C.,  {Sanders} J.~S.,  {Taylor} G.~B.,  {Allen} S.~W.,  {Crawford}
  C.~S.,  {Johnstone} R.~M.,   {Iwasawa} K.,  2006, \mn@doi [\mnras]
  {10.1111/j.1365-2966.2005.09896.x}, \href
  {http://adsabs.harvard.edu/abs/2006MNRAS.366..417F} {366, 417}

\bibitem[\protect\citeauthoryear{{Fabian}, {Walker}, {Russell}, {Pinto},
  {Sanders}  \& {Reynolds}}{{Fabian} et~al.}{2017}]{Fabian2017}
{Fabian} A.~C.,  {Walker} S.~A.,  {Russell} H.~R.,  {Pinto} C.,  {Sanders}
  J.~S.,   {Reynolds} C.~S.,  2017, \mn@doi [\mnras] {10.1093/mnrasl/slw170},
  \href {http://adsabs.harvard.edu/abs/2017MNRAS.464L...1F} {464, L1}

\bibitem[\protect\citeauthoryear{{Feretti}, {Giovannini}, {Govoni}  \&
  {Murgia}}{{Feretti} et~al.}{2012}]{Feretti2012}
{Feretti} L.,  {Giovannini} G.,  {Govoni} F.,   {Murgia} M.,  2012, \mn@doi
  [\aapr] {10.1007/s00159-012-0054-z}, \href
  {http://adsabs.harvard.edu/abs/2012A%26ARv..20...54F} {20, 54}

\bibitem[\protect\citeauthoryear{{Fujita} \& {Ohira}}{{Fujita} \&
  {Ohira}}{2011}]{Fujita2011}
{Fujita} Y.,  {Ohira} Y.,  2011, \mn@doi [\apj] {10.1088/0004-637X/738/2/182},
  \href {http://adsabs.harvard.edu/abs/2011ApJ...738..182F} {738, 182}

\bibitem[\protect\citeauthoryear{{Fujita} \& {Ohira}}{{Fujita} \&
  {Ohira}}{2012}]{Fujita2012}
{Fujita} Y.,  {Ohira} Y.,  2012, \mn@doi [\apj] {10.1088/0004-637X/746/1/53},
  \href {http://adsabs.harvard.edu/abs/2012ApJ...746...53F} {746, 53}

\bibitem[\protect\citeauthoryear{{Fujita} \& {Ohira}}{{Fujita} \&
  {Ohira}}{2013}]{Fujita2013}
{Fujita} Y.,  {Ohira} Y.,  2013, \mn@doi [\mnras] {10.1093/mnras/sts050}, \href
  {http://adsabs.harvard.edu/abs/2013MNRAS.428..599F} {428, 599}

\bibitem[\protect\citeauthoryear{{Gaspari}, {Brighenti}  \& {Temi}}{{Gaspari}
  et~al.}{2012}]{2012MNRAS.424..190G}
{Gaspari} M.,  {Brighenti} F.,   {Temi} P.,  2012, \mn@doi [\mnras]
  {10.1111/j.1365-2966.2012.21183.x}, \href
  {http://adsabs.harvard.edu/abs/2012MNRAS.424..190G} {424, 190}

\bibitem[\protect\citeauthoryear{{Giacintucci}, {Markevitch}, {Venturi},
  {Clarke}, {Cassano}  \& {Mazzotta}}{{Giacintucci}
  et~al.}{2014}]{Giacintucci2014}
{Giacintucci} S.,  {Markevitch} M.,  {Venturi} T.,  {Clarke} T.~E.,  {Cassano}
  R.,   {Mazzotta} P.,  2014, \mn@doi [\apj] {10.1088/0004-637X/781/1/9}, \href
  {http://adsabs.harvard.edu/abs/2014ApJ...781....9G} {781, 9}

\bibitem[\protect\citeauthoryear{{Guo} \& {Mathews}}{{Guo} \&
  {Mathews}}{2010}]{Guo2010b}
{Guo} F.,  {Mathews} W.~G.,  2010, \mn@doi [\apj]
  {10.1088/0004-637X/717/2/937}, \href
  {http://adsabs.harvard.edu/abs/2010ApJ...717..937G} {717, 937}

\bibitem[\protect\citeauthoryear{{Guo} \& {Mathews}}{{Guo} \&
  {Mathews}}{2011}]{Guo2011}
{Guo} F.,  {Mathews} W.~G.,  2011, \mn@doi [\apj]
  {10.1088/0004-637X/728/2/121}, \href
  {http://adsabs.harvard.edu/abs/2011ApJ...728..121G} {728, 121}

\bibitem[\protect\citeauthoryear{{Guo} \& {Oh}}{{Guo} \& {Oh}}{2008}]{Guo2008}
{Guo} F.,  {Oh} S.~P.,  2008, \mn@doi [\mnras]
  {10.1111/j.1365-2966.2007.12692.x}, \href
  {http://adsabs.harvard.edu/abs/2008MNRAS.384..251G} {384, 251}

\bibitem[\protect\citeauthoryear{{Guo} \& {Oh}}{{Guo} \& {Oh}}{2009}]{Guo2009}
{Guo} F.,  {Oh} S.~P.,  2009, \mn@doi [\mnras]
  {10.1111/j.1365-2966.2009.15592.x}, \href
  {http://adsabs.harvard.edu/abs/2009MNRAS.400.1992G} {400, 1992}

\bibitem[\protect\citeauthoryear{{Hillel} \& {Soker}}{{Hillel} \&
  {Soker}}{2016}]{Hillel2016}
{Hillel} S.,  {Soker} N.,  2016, \mn@doi [\mnras] {10.1093/mnras/stv2483},
  \href {http://adsabs.harvard.edu/abs/2016MNRAS.455.2139H} {455, 2139}

\bibitem[\protect\citeauthoryear{{Hitomi Collaboration} et~al.,}{{Hitomi
  Collaboration} et~al.}{2016}]{Hitomi2016}
{Hitomi Collaboration} et~al., 2016, \mn@doi [\nat] {10.1038/nature18627},
  \href {http://adsabs.harvard.edu/abs/2016Natur.535..117H} {535, 117}

\bibitem[\protect\citeauthoryear{{Hoffer}, {Donahue}, {Hicks}  \&
  {Barthelemy}}{{Hoffer} et~al.}{2012}]{Hoffer2012}
{Hoffer} A.~S.,  {Donahue} M.,  {Hicks} A.,   {Barthelemy} R.~S.,  2012,
  \mn@doi [\apjs] {10.1088/0067-0049/199/1/23}, \href
  {http://adsabs.harvard.edu/abs/2012ApJS..199...23H} {199, 23}

\bibitem[\protect\citeauthoryear{{Hudson}, {Mittal}, {Reiprich}, {Nulsen},
  {Andernach}  \& {Sarazin}}{{Hudson} et~al.}{2010}]{Hudson2010}
{Hudson} D.~S.,  {Mittal} R.,  {Reiprich} T.~H.,  {Nulsen} P.~E.~J.,
  {Andernach} H.,   {Sarazin} C.~L.,  2010, \mn@doi [\aap]
  {10.1051/0004-6361/200912377}, \href
  {http://adsabs.harvard.edu/abs/2010A%26A...513A..37H} {513, A37}

\bibitem[\protect\citeauthoryear{{Jacob} \& {Pfrommer}}{{Jacob} \&
  {Pfrommer}}{2017}]{Jacob2016a}
{Jacob} S.,  {Pfrommer} C.,  2017, \mn@doi [\mnras] {10.1093/mnras/stx131},
  \href {http://adsabs.harvard.edu/abs/2017MNRAS.467.1449J} {467, 1449}

\bibitem[\protect\citeauthoryear{{Kannan}, {Vogelsberger}, {Pfrommer},
  {Weinberger}, {Springel}, {Hernquist}, {Puchwein}  \& {Pakmor}}{{Kannan}
  et~al.}{2016}]{Kannan2016}
{Kannan} R.,  {Vogelsberger} M.,  {Pfrommer} C.,  {Weinberger} R.,  {Springel}
  V.,  {Hernquist} L.,  {Puchwein} E.,   {Pakmor} R.,  2016, preprint, \href
  {http://adsabs.harvard.edu/abs/2016arXiv161201522K} {} (\mn@eprint {arXiv}
  {1612.01522})

\bibitem[\protect\citeauthoryear{{Kim} \& {Narayan}}{{Kim} \&
  {Narayan}}{2003}]{Kim2003Turb}
{Kim} W.-T.,  {Narayan} R.,  2003, \mn@doi [\apjl] {10.1086/379342}, \href
  {http://adsabs.harvard.edu/abs/2003ApJ...596L.139K} {596, L139}

\bibitem[\protect\citeauthoryear{{Kuchar} \& {En{\ss}lin}}{{Kuchar} \&
  {En{\ss}lin}}{2011}]{Kuchar2011}
{Kuchar} P.,  {En{\ss}lin} T.~A.,  2011, \mn@doi [\aap]
  {10.1051/0004-6361/200913918}, \href
  {http://adsabs.harvard.edu/abs/2011A%26A...529A..13K} {529, A13}

\bibitem[\protect\citeauthoryear{{Kulsrud} \& {Pearce}}{{Kulsrud} \&
  {Pearce}}{1969}]{Kulsrud1969}
{Kulsrud} R.,  {Pearce} W.~P.,  1969, \mn@doi [\apj] {10.1086/149981}, \href
  {http://adsabs.harvard.edu/abs/1969ApJ...156..445K} {156, 445}

\bibitem[\protect\citeauthoryear{{Lagan{\'a}}, {Martinet}, {Durret}, {Lima
  Neto}, {Maughan}  \& {Zhang}}{{Lagan{\'a}} et~al.}{2013}]{Lagana2013}
{Lagan{\'a}} T.~F.,  {Martinet} N.,  {Durret} F.,  {Lima Neto} G.~B.,
  {Maughan} B.,   {Zhang} Y.-Y.,  2013, \mn@doi [\aap]
  {10.1051/0004-6361/201220423}, \href
  {http://adsabs.harvard.edu/abs/2013A%26A...555A..66L} {555, A66}

\bibitem[\protect\citeauthoryear{{Landry}, {Bonamente}, {Giles}, {Maughan},
  {Joy}  \& {Murray}}{{Landry} et~al.}{2013}]{Landry2013}
{Landry} D.,  {Bonamente} M.,  {Giles} P.,  {Maughan} B.,  {Joy} M.,   {Murray}
  S.,  2013, \mn@doi [\mnras] {10.1093/mnras/stt901}, \href
  {http://adsabs.harvard.edu/abs/2013MNRAS.433.2790L} {433, 2790}

\bibitem[\protect\citeauthoryear{{Loewenstein}, {Zweibel}  \&
  {Begelman}}{{Loewenstein} et~al.}{1991}]{Loewenstein1991}
{Loewenstein} M.,  {Zweibel} E.~G.,   {Begelman} M.~C.,  1991, \mn@doi [\apj]
  {10.1086/170369}, \href {http://adsabs.harvard.edu/abs/1991ApJ...377..392L}
  {377, 392}

\bibitem[\protect\citeauthoryear{{Markevitch} \& {Vikhlinin}}{{Markevitch} \&
  {Vikhlinin}}{2007}]{Markevitch2007}
{Markevitch} M.,  {Vikhlinin} A.,  2007, \mn@doi [\physrep]
  {10.1016/j.physrep.2007.01.001}, \href
  {http://adsabs.harvard.edu/abs/2007PhR...443....1M} {443, 1}

\bibitem[\protect\citeauthoryear{{McNamara}, {Nulsen}, {Wise}, {Rafferty},
  {Carilli}, {Sarazin}  \& {Blanton}}{{McNamara}
  et~al.}{2005}]{2005Natur.433...45M}
{McNamara} B.~R.,  {Nulsen} P.~E.~J.,  {Wise} M.~W.,  {Rafferty} D.~A.,
  {Carilli} C.,  {Sarazin} C.~L.,   {Blanton} E.~L.,  2005, \mn@doi [\nat]
  {10.1038/nature03202}, \href
  {http://adsabs.harvard.edu/abs/2005Natur.433...45M} {433, 45}

\bibitem[\protect\citeauthoryear{{Murgia}, {Govoni}, {Markevitch}, {Feretti},
  {Giovannini}, {Taylor}  \& {Carretti}}{{Murgia} et~al.}{2009}]{Murgia2009}
{Murgia} M.,  {Govoni} F.,  {Markevitch} M.,  {Feretti} L.,  {Giovannini} G.,
  {Taylor} G.~B.,   {Carretti} E.,  2009, \mn@doi [\aap]
  {10.1051/0004-6361/200911659}, \href
  {http://adsabs.harvard.edu/abs/2009A%26A...499..679M} {499, 679}

\bibitem[\protect\citeauthoryear{{Murgia}, {Eckert}, {Govoni}, {Ferrari},
  {Pandey-Pommier}, {Nevalainen}  \& {Paltani}}{{Murgia}
  et~al.}{2010}]{Murgia2010}
{Murgia} M.,  {Eckert} D.,  {Govoni} F.,  {Ferrari} C.,  {Pandey-Pommier} M.,
  {Nevalainen} J.,   {Paltani} S.,  2010, \mn@doi [\aap]
  {10.1051/0004-6361/201014126}, \href
  {http://adsabs.harvard.edu/abs/2010A%26A...514A..76M} {514, A76}

\bibitem[\protect\citeauthoryear{{Nulsen} \& {B{\"o}hringer}}{{Nulsen} \&
  {B{\"o}hringer}}{1995}]{Nulsen1995}
{Nulsen} P.~E.~J.,  {B{\"o}hringer} H.,  1995, \mn@doi [\mnras]
  {10.1093/mnras/274.4.1093}, \href
  {http://adsabs.harvard.edu/abs/1995MNRAS.274.1093N} {274, 1093}

\bibitem[\protect\citeauthoryear{{Nulsen}, {McNamara}, {Wise}  \&
  {David}}{{Nulsen} et~al.}{2005}]{2005ApJ...628..629N}
{Nulsen} P.~E.~J.,  {McNamara} B.~R.,  {Wise} M.~W.,   {David} L.~P.,  2005,
  \mn@doi [\apj] {10.1086/430845}, \href
  {http://adsabs.harvard.edu/abs/2005ApJ...628..629N} {628, 629}

\bibitem[\protect\citeauthoryear{{O'Dea} et~al.,}{{O'Dea}
  et~al.}{2008}]{ODea2008}
{O'Dea} C.~P.,  et~al., 2008, \mn@doi [\apj] {10.1086/588212}, \href
  {http://adsabs.harvard.edu/abs/2008ApJ...681.1035O} {681, 1035}

\bibitem[\protect\citeauthoryear{{Ogrean}, {Hatch}, {Simionescu},
  {B{\"o}hringer}, {Br{\"u}ggen}, {Fabian}  \& {Werner}}{{Ogrean}
  et~al.}{2010}]{Ogrean2010}
{Ogrean} G.~A.,  {Hatch} N.~A.,  {Simionescu} A.,  {B{\"o}hringer} H.,
  {Br{\"u}ggen} M.,  {Fabian} A.~C.,   {Werner} N.,  2010, \mn@doi [\mnras]
  {10.1111/j.1365-2966.2010.16718.x}, \href
  {http://adsabs.harvard.edu/abs/2010MNRAS.406..354O} {406, 354}

\bibitem[\protect\citeauthoryear{{Peterson} \& {Fabian}}{{Peterson} \&
  {Fabian}}{2006}]{Peterson2006}
{Peterson} J.~R.,  {Fabian} A.~C.,  2006, \mn@doi [\physrep]
  {10.1016/j.physrep.2005.12.007}, \href
  {http://adsabs.harvard.edu/abs/2006PhR...427....1P} {427, 1}

\bibitem[\protect\citeauthoryear{{Pfrommer}}{{Pfrommer}}{2008}]{Pfrommer2008_III}
{Pfrommer} C.,  2008, \mn@doi [\mnras] {10.1111/j.1365-2966.2008.12957.x},
  \href {http://adsabs.harvard.edu/abs/2008MNRAS.385.1242P} {385, 1242}

\bibitem[\protect\citeauthoryear{{Pfrommer}}{{Pfrommer}}{2013}]{Pfrommer2013}
{Pfrommer} C.,  2013, \mn@doi [\apj] {10.1088/0004-637X/779/1/10}, \href
  {http://adsabs.harvard.edu/abs/2013ApJ...779...10P} {779, 10}

\bibitem[\protect\citeauthoryear{{Pfrommer} \& {En{\ss}lin}}{{Pfrommer} \&
  {En{\ss}lin}}{2004a}]{Pfrommer2004c}
{Pfrommer} C.,  {En{\ss}lin} T.~A.,  2004a, \mn@doi [\mnras]
  {10.1111/j.1365-2966.2004.07900.x}, \href
  {http://adsabs.harvard.edu/abs/2004MNRAS.352...76P} {352, 76}

\bibitem[\protect\citeauthoryear{{Pfrommer} \& {En{\ss}lin}}{{Pfrommer} \&
  {En{\ss}lin}}{2004b}]{Pfrommer2004}
{Pfrommer} C.,  {En{\ss}lin} T.~A.,  2004b, \mn@doi [\aap]
  {10.1051/0004-6361:20031464}, \href
  {http://adsabs.harvard.edu/abs/2004A%26A...413...17P} {413, 17}

\bibitem[\protect\citeauthoryear{{Pfrommer}, {En{\ss}lin}  \&
  {Springel}}{{Pfrommer} et~al.}{2008}]{Pfrommer2008}
{Pfrommer} C.,  {En{\ss}lin} T.~A.,   {Springel} V.,  2008, \mn@doi [\mnras]
  {10.1111/j.1365-2966.2008.12956.x}, \href
  {http://adsabs.harvard.edu/abs/2008MNRAS.385.1211P} {385, 1211}

\bibitem[\protect\citeauthoryear{{Pfrommer}, {Chang}  \&
  {Broderick}}{{Pfrommer} et~al.}{2012}]{2012ApJ...752...24P}
{Pfrommer} C.,  {Chang} P.,   {Broderick} A.~E.,  2012, \mn@doi [\apj]
  {10.1088/0004-637X/752/1/24}, \href
  {http://adsabs.harvard.edu/abs/2012ApJ...752...24P} {752, 24}

\bibitem[\protect\citeauthoryear{{Pfrommer}, {Pakmor}, {Schaal}, {Simpson}  \&
  {Springel}}{{Pfrommer} et~al.}{2017}]{Pfrommer2016}
{Pfrommer} C.,  {Pakmor} R.,  {Schaal} K.,  {Simpson} C.~M.,   {Springel} V.,
  2017, \mn@doi [\mnras] {10.1093/mnras/stw2941}, \href
  {http://adsabs.harvard.edu/abs/2017MNRAS.465.4500P} {465, 4500}

\bibitem[\protect\citeauthoryear{{Pinzke} \& {Pfrommer}}{{Pinzke} \&
  {Pfrommer}}{2010}]{Pinzke2010}
{Pinzke} A.,  {Pfrommer} C.,  2010, \mn@doi [\mnras]
  {10.1111/j.1365-2966.2010.17328.x}, \href
  {http://adsabs.harvard.edu/abs/2010MNRAS.409..449P} {409, 449}

\bibitem[\protect\citeauthoryear{{Pinzke}, {Pfrommer}  \&
  {Bergstr{\"o}m}}{{Pinzke} et~al.}{2011}]{Pinzke2011}
{Pinzke} A.,  {Pfrommer} C.,   {Bergstr{\"o}m} L.,  2011, \mn@doi [\prd]
  {10.1103/PhysRevD.84.123509}, \href
  {http://adsabs.harvard.edu/abs/2011PhRvD..84l3509P} {84, 123509}

\bibitem[\protect\citeauthoryear{{Reynolds}, {Balbus}  \&
  {Schekochihin}}{{Reynolds} et~al.}{2015}]{Reynolds2015}
{Reynolds} C.~S.,  {Balbus} S.~A.,   {Schekochihin} A.~A.,  2015, \mn@doi
  [\apj] {10.1088/0004-637X/815/1/41}, \href
  {http://adsabs.harvard.edu/abs/2015ApJ...815...41R} {815, 41}

\bibitem[\protect\citeauthoryear{{Ruszkowski} \& {Begelman}}{{Ruszkowski} \&
  {Begelman}}{2002}]{2002ApJ...581..223R}
{Ruszkowski} M.,  {Begelman} M.~C.,  2002, \mn@doi [\apj] {10.1086/344170},
  \href {http://adsabs.harvard.edu/abs/2002ApJ...581..223R} {581, 223}

\bibitem[\protect\citeauthoryear{{Sarazin}}{{Sarazin}}{1999}]{Sarazin1999}
{Sarazin} C.~L.,  1999, \mn@doi [\apj] {10.1086/307501}, \href
  {http://adsabs.harvard.edu/abs/1999ApJ...520..529S} {520, 529}

\bibitem[\protect\citeauthoryear{{Sayers} et~al.,}{{Sayers}
  et~al.}{2013}]{Sayers2013}
{Sayers} J.,  et~al., 2013, \mn@doi [\apj] {10.1088/0004-637X/764/2/152}, \href
  {http://adsabs.harvard.edu/abs/2013ApJ...764..152S} {764, 152}

\bibitem[\protect\citeauthoryear{{Shabala}, {Ash}, {Alexander}  \&
  {Riley}}{{Shabala} et~al.}{2008}]{2008MNRAS.388..625S}
{Shabala} S.~S.,  {Ash} S.,  {Alexander} P.,   {Riley} J.~M.,  2008, \mn@doi
  [\mnras] {10.1111/j.1365-2966.2008.13459.x}, \href
  {http://adsabs.harvard.edu/abs/2008MNRAS.388..625S} {388, 625}

\bibitem[\protect\citeauthoryear{{Sharma}, {Chandran}, {Quataert}  \&
  {Parrish}}{{Sharma} et~al.}{2009}]{2009ApJ...699..348S}
{Sharma} P.,  {Chandran} B.~D.~G.,  {Quataert} E.,   {Parrish} I.~J.,  2009,
  \mn@doi [\apj] {10.1088/0004-637X/699/1/348}, \href
  {http://adsabs.harvard.edu/abs/2009ApJ...699..348S} {699, 348}

\bibitem[\protect\citeauthoryear{{Sijacki}, {Pfrommer}, {Springel}  \&
  {En{\ss}lin}}{{Sijacki} et~al.}{2008}]{Sijacki2008}
{Sijacki} D.,  {Pfrommer} C.,  {Springel} V.,   {En{\ss}lin} T.~A.,  2008,
  \mn@doi [\mnras] {10.1111/j.1365-2966.2008.13310.x}, \href
  {http://adsabs.harvard.edu/abs/2008MNRAS.387.1403S} {387, 1403}

\bibitem[\protect\citeauthoryear{{Sijbring}}{{Sijbring}}{1993}]{Sijbring1993}
{Sijbring} L.~G.,  1993, {A radio continuum and HI line study of the perseus
  cluster}

\bibitem[\protect\citeauthoryear{{Simionescu} et~al.,}{{Simionescu}
  et~al.}{2012}]{Simionescu2012}
{Simionescu} A.,  et~al., 2012, \mn@doi [\apj] {10.1088/0004-637X/757/2/182},
  \href {http://adsabs.harvard.edu/abs/2012ApJ...757..182S} {757, 182}

\bibitem[\protect\citeauthoryear{{Urban}, {Werner}, {Simionescu}, {Allen}  \&
  {B{\"o}hringer}}{{Urban} et~al.}{2011}]{Urban2011}
{Urban} O.,  {Werner} N.,  {Simionescu} A.,  {Allen} S.~W.,   {B{\"o}hringer}
  H.,  2011, \mn@doi [\mnras] {10.1111/j.1365-2966.2011.18526.x}, \href
  {http://adsabs.harvard.edu/abs/2011MNRAS.414.2101U} {414, 2101}

\bibitem[\protect\citeauthoryear{{Vikhlinin}, {Kravtsov}, {Forman}, {Jones},
  {Markevitch}, {Murray}  \& {Van Speybroeck}}{{Vikhlinin}
  et~al.}{2006}]{Vikhlinin2006}
{Vikhlinin} A.,  {Kravtsov} A.,  {Forman} W.,  {Jones} C.,  {Markevitch} M.,
  {Murray} S.~S.,   {Van Speybroeck} L.,  2006, \mn@doi [\apj]
  {10.1086/500288}, \href {http://adsabs.harvard.edu/abs/2006ApJ...640..691V}
  {640, 691}

\bibitem[\protect\citeauthoryear{{Vogt} \& {En{\ss}lin}}{{Vogt} \&
  {En{\ss}lin}}{2005}]{Vogt2005}
{Vogt} C.,  {En{\ss}lin} T.~A.,  2005, \mn@doi [\aap]
  {10.1051/0004-6361:20041839}, \href
  {http://adsabs.harvard.edu/abs/2005A%26A...434...67V} {434, 67}

\bibitem[\protect\citeauthoryear{{Voit}, {Cavagnolo}, {Donahue}, {Rafferty},
  {McNamara}  \& {Nulsen}}{{Voit} et~al.}{2008}]{Voit2008}
{Voit} G.~M.,  {Cavagnolo} K.~W.,  {Donahue} M.,  {Rafferty} D.~A.,  {McNamara}
  B.~R.,   {Nulsen} P.~E.~J.,  2008, \mn@doi [\apjl] {10.1086/590344}, \href
  {http://adsabs.harvard.edu/abs/2008ApJ...681L...5V} {681, L5}

\bibitem[\protect\citeauthoryear{{Wentzel}}{{Wentzel}}{1971}]{Wentzel1971}
{Wentzel} D.~G.,  1971, \mn@doi [\apj] {10.1086/150794}, \href
  {http://adsabs.harvard.edu/abs/1971ApJ...163..503W} {163, 503}

\bibitem[\protect\citeauthoryear{{Wiener}, {Oh}  \& {Guo}}{{Wiener}
  et~al.}{2013}]{Wiener2013}
{Wiener} J.,  {Oh} S.~P.,   {Guo} F.,  2013, \mn@doi [\mnras]
  {10.1093/mnras/stt1163}, \href
  {http://adsabs.harvard.edu/abs/2013MNRAS.434.2209W} {434, 2209}

\bibitem[\protect\citeauthoryear{{Yang} \& {Reynolds}}{{Yang} \&
  {Reynolds}}{2016a}]{Yang2016a}
{Yang} H.-Y.~K.,  {Reynolds} C.~S.,  2016a, \mn@doi [\apj]
  {10.3847/0004-637X/818/2/181}, \href
  {http://adsabs.harvard.edu/abs/2016ApJ...818..181Y} {818, 181}

\bibitem[\protect\citeauthoryear{{Yang} \& {Reynolds}}{{Yang} \&
  {Reynolds}}{2016b}]{Yang2016b}
{Yang} H.-Y.~K.,  {Reynolds} C.~S.,  2016b, \mn@doi [\apj]
  {10.3847/0004-637X/829/2/90}, \href
  {http://adsabs.harvard.edu/abs/2016ApJ...829...90Y} {829, 90}

\bibitem[\protect\citeauthoryear{{Zandanel}, {Pfrommer}  \& {Prada}}{{Zandanel}
  et~al.}{2014}]{Zandanel2014}
{Zandanel} F.,  {Pfrommer} C.,   {Prada} F.,  2014, \mn@doi [\mnras]
  {10.1093/mnras/stt2250}, \href
  {http://adsabs.harvard.edu/abs/2014MNRAS.438..124Z} {438, 124}

\bibitem[\protect\citeauthoryear{{Zhuravleva} et~al.,}{{Zhuravleva}
  et~al.}{2014}]{Zhuravleva2014}
{Zhuravleva} I.,  et~al., 2014, \mn@doi [\nat] {10.1038/nature13830}, \href
  {http://adsabs.harvard.edu/abs/2014Natur.515...85Z} {515, 85}

\bibitem[\protect\citeauthoryear{{ZuHone}, {Markevitch}, {Ruszkowski}  \&
  {Lee}}{{ZuHone} et~al.}{2013}]{ZuHone2013}
{ZuHone} J.~A.,  {Markevitch} M.,  {Ruszkowski} M.,   {Lee} D.,  2013, \mn@doi
  [\apj] {10.1088/0004-637X/762/2/69}, \href
  {http://adsabs.harvard.edu/abs/2013ApJ...762...69Z} {762, 69}

\bibitem[\protect\citeauthoryear{{de Gasperin} et~al.,}{{de Gasperin}
  et~al.}{2012}]{deGasperin2012}
{de Gasperin} F.,  et~al., 2012, \mn@doi [\aap] {10.1051/0004-6361/201220209},
  \href {http://adsabs.harvard.edu/abs/2012A%26A...547A..56D} {547, A56}

\makeatother
\end{thebibliography}

%%%%%%%%%%%%%%%%%%%%%%%%%%%%%%%%%%%%%%%%%%%%%%%%%%

%%%%%%%%%%%%%%%%% APPENDICES %%%%%%%%%%%%%%%%%%%%%

\appendix

\section{Density fits}
\label{sec:appdensity}

In Table~\ref{tab:fitparam} we list the fit parameters of the density profile for the 24 clusters for which we performed our own fits.

\begin{table}
\caption{Parameters for density profiles.}
\label{tab:fitparam}
\begin{threeparttable}
\begin{tabular}{l r r r r}
\hline
Cluster 	&\multicolumn{1}{c}{ $r_{\rmn{cut,}n_\rmn{e}}^{\rmn{(1)}}$} & \multicolumn{1}{c}{$n_0$} &\multicolumn{1}{c}{ $\beta$} & \multicolumn{1}{c}{$r_\rmn{c}$} \\
		&\multicolumn{1}{c}{ (kpc) }&\multicolumn{1}{c}{ ($\rmn{cm}^{-3}$)}& &\multicolumn{1}{c}{(kpc)}\\
\hline
Centaurus	&	62	&	0.225	&	0.30	&	0.9\hphantom{$^{\rmn{(1)}}$}	\\
Hydra A	&	296		&	0.067	&	0.40	&	11.2\hphantom{$^{\rmn{(1)}}$}	\\
Virgo	&	44		&	0.230	&	0.29	&	0.6\hphantom{$^{\rmn{(1)}}$}	\\
A 85	&	248		&	0.089	&	0.34	&	7.2\hphantom{$^{\rmn{(1)}}$}	\\
A 496	&	79	&	0.088	&	0.32	&	4.9\hphantom{$^{\rmn{(1)}}$}	\\
A 539	&	311	&	0.068	&	0.24	&	0.5$^{\rmn{(2)}}$	\\
A 1644	&	284	&	0.051	&	0.26	&	2.1\hphantom{$^{\rmn{(1)}}$}	\\
A 2052	&	112	&	0.027	&	0.41	&	18.7\hphantom{$^{\rmn{(1)}}$}	\\
A 2199	&	84	&	0.101	&	0.25	&	2.2\hphantom{$^{\rmn{(1)}}$}	\\
A 2597	&	87	&	0.083	&	0.43	&	17.0\hphantom{$^{\rmn{(1)}}$}	\\
A 3112	&	226	&	0.079	&	0.40	&	10.2\hphantom{$^{\rmn{(1)}}$}	\\
A 3581	&	105	&	0.043	&	0.39	&	6.9\hphantom{$^{\rmn{(1)}}$}	\\
A 4059	&	213	&	0.053	&	0.29	&	3.9\hphantom{$^{\rmn{(1)}}$}	\\
AWM 7	&	78	&	0.113	&	0.22	&	0.5$^{\rmn{(2)}}$	\\
MKW 3S	&	386	&	0.027	&	0.45	&	21.9\hphantom{$^{\rmn{(1)}}$}	\\
PKS 0745	&	496	&	0.112	&	0.52	&	28.0\hphantom{$^{\rmn{(1)}}$}	\\
ZwCl 1742	&	343	&	0.029	&	0.56	&	30.3\hphantom{$^{\rmn{(1)}}$}	\\
Ophiuchus	&	257	&	0.463	&	0.26	&	0.5$^{\rmn{(2)}}$	\\
Perseus	&	114	&	0.049	&	0.62	&	42.4\hphantom{$^{\rmn{(1)}}$}	\\
2A 0335	&	148	&	0.095	&	0.45	&	12.0\hphantom{$^{\rmn{(1)}}$}	\\
RBS 797	&	537	&	0.101	&	0.65	&	43.2\hphantom{$^{\rmn{(1)}}$}\\
RX J1347	&	988	&	0.103	&	0.65	&	54.3\hphantom{$^{\rmn{(1)}}$}	\\
RX J1504	&	587	&	0.163	&	0.62	&	31.8\hphantom{$^{\rmn{(1)}}$}	\\
RX J1532	&	477	&	0.091	&	0.62	&	38.9\hphantom{$^{\rmn{(1)}}$}	\\
\hline
\end{tabular}
\end{threeparttable}
\begin{tablenotes}
\item (1) Maximal radius that we include in fit.
\item (2) Parameter fixed during the fit.
\end{tablenotes}
\end{table}

\section{Calculate non-thermal emission}
\label{sec:appemission}

\subsection{Radio emission}
\label{sec:appradio}

We calculate the radio emission from secondary electrons following \citet{Pfrommer2008}. Therefore, we model the CR proton population as
\begin{equation}
f_\rmn{p} (p_\rmn{p}) = \frac{\rmn{d} N}{\rmn{d} p_\rmn{p} \rmn{d} V} = C_\rmn{p} p_\rmn{p}^{-\alpha_\rmn{p}} \theta(p_\rmn{p}-q_\rmn{p})\label{eq:CRpop}
\end{equation}
where $p_\rmn{p} = P_\rmn{p}/(m_\rmn{p} c)$ is the dimensionless proton momentum. The CR spectrum is a power law in momentum with spectral index $\alpha_\rmn{p}=2.4$.
$\theta(x)$ denotes the Heaviside step function, which imposes a lower momentum cut-off at $q_\rmn{p}=0.5$. $C_\rmn{p}$ describes the normalization, which we obtain from the CR pressure as \citep{Ensslin2007, Pfrommer2008}
\begin{equation}
C_\rmn{p} (r) = \frac{6 P_\rmn{cr, ex}(r)}{m_\rmn{p} c^2} \left[\mathcal{B}_{\frac{1}{1+q_\rmn{p}^2}}\left(\frac{\alpha_\rmn{p}-2}{2}, \frac{3 - \alpha_\rmn{p}}{2}\right)\right]^{-1}.
\end{equation}
Here, $\mathcal{B}_x(a,b)$ denotes the incomplete beta function and we use the
extrapolated CR pressure profile, $P_\rmn{cr,ex}(r)$, from the steady state solutions (see
Equation~\ref{eq:Pex}).

This CR population interacts hadronically with the nucleons in the ICM and produces pions. The charged pions decay into muons and eventually into electrons. The resulting electron distribution in steady state, where hadronic injection balances radiative cooling, can again be described by a power-law spectrum in momentum with spectral index $\alpha_\rmn{e} = \alpha_\rmn{p}+1$. 
The resulting synchrotron emissivity at frequency $\nu$ per steradian by a secondary population of CR electrons with an isotropic distribution of pitch angles is given by
\begin{equation}
  \label{eq:jnu2}
  j_\nu = \frac{A_\nu}{4 \upi} C_\rmn{p} n_\rmn{N} \frac{e_B}{e_B + e_\rmn{rad}}
  \left(\frac{e_B}{e_{B_\rmn{c}}}\right)^{\left(\alpha_\nu -1\right)/2}
\end{equation}
where $C_\rmn{p}$ is the normalization of the proton spectrum and $n_\rmn{N} = n_\rmn{H} + 4 n_\rmn{He} = \rho/m_\rmn{p}$ the nucleon number density. The nucleon number density is related to the electron number density by $n_\rmn{N} = \mu_\rmn{e} n_\rmn{e}$. We use our fits to the ACCEPT data to describe $n_\rmn{e}$ (see Section~\ref{sec:density}) and a mean molecular weight per electron of $\mu_\rmn{e} = 1.18$. This corresponds to a composition of the ICM with hydrogen mass fraction $X = 0.7$ and helium mass fraction $Y = 0.28$.

The quantities $e_B$, $e_\rmn{rad}$ and $e_{B_\rmn{c}}$ describe the energy
densities that are relevant for synchrotron radiation.  $e_B = B^2/(8 \upi)$
denotes the magnetic energy density that we parametrize in
equation~\ref{eq:Bfield}.  Inverse Compton scattering on radiation fields cools
the electron population and is thus important for determining its steady state
distribution. The total radiation field in galaxy clusters is composed of CMB
photons and the emission from dust and stars such that $e_\rmn{rad} =
e_{\rmn{CMB}} + e_\rmn{SD}$. Here, we treat the energy density of CMB photons
with an equivalent magnetic field of $B_\rmn{CMB} = 3.24 (1+z)^2\rmn{~\umu G}$
\citep{Pfrommer2008}. For the emission from stars and dust, we employ the model
by \citet{Pinzke2011}.\footnote{We are correcting two typos in equations (A8) and (A9) of
    \citet{Pinzke2011} and replace the factors $6.0 \times 10^{-9}$ and $4.0\times 10^{-7}\rmn{~kpc^2}$ by $71$ and $4384\rmn{~kpc^2}$, respectively, so that we can reproduce
    the correct results in fig. 22 of \citet{Pinzke2011}.}

\begin{figure*}
  \includegraphics{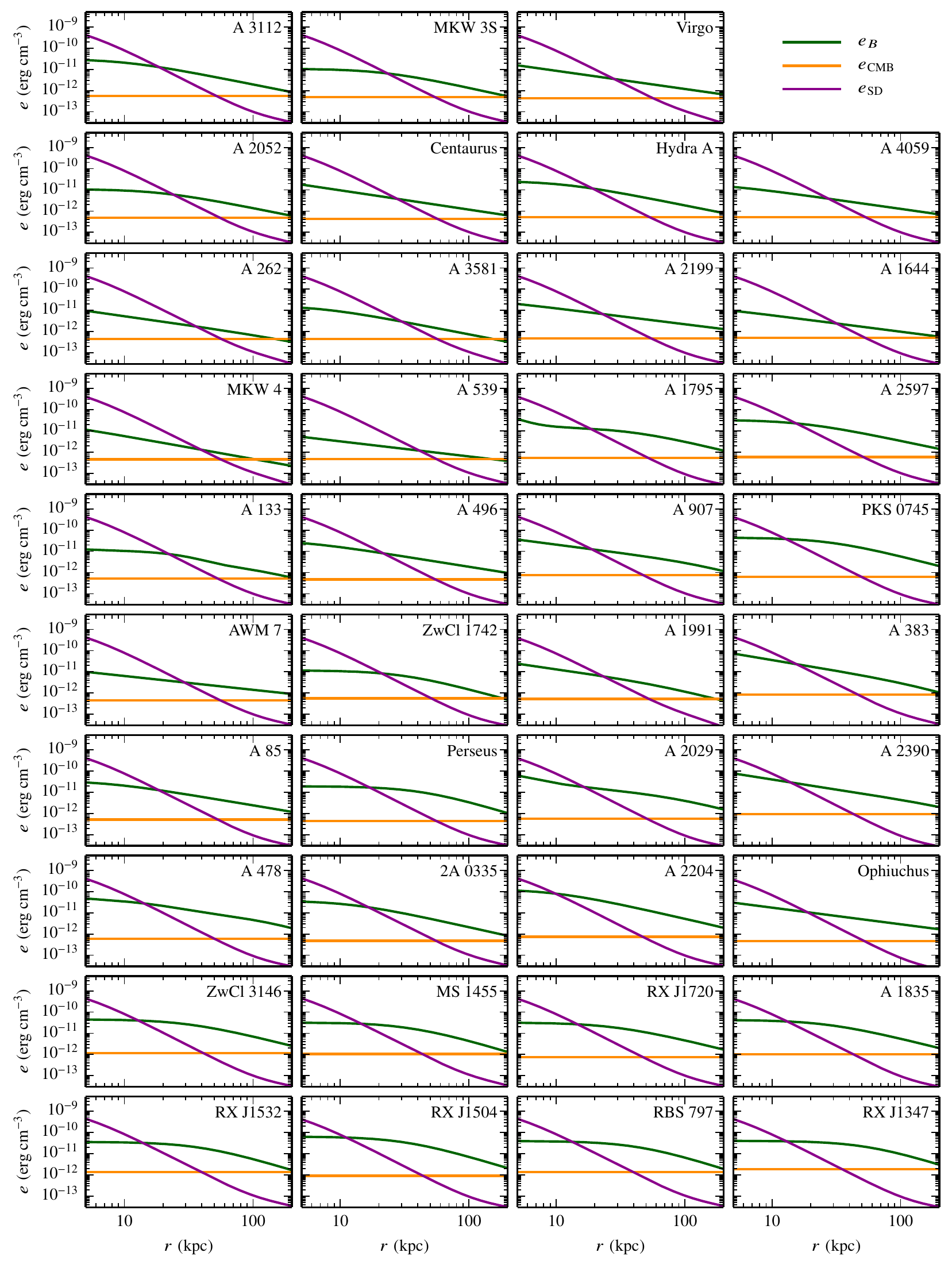}
  \caption{Energy density profiles of the magnetic field, the CMB, and
    the radiation field emitted by stars and dust. The clusters are ordered
    by row, starting with non-RMH clusters and followed by RMH-hosting clusters
    from Perseus onward.}
    \label{fig:energydensities}
\end{figure*}

In Fig.~\ref{fig:energydensities} we show radial profiles of $e_B$,
$e_{\rmn{CMB}}$ and $e_{\rmn{SD}}$ for our entire cluster sample. At small
radii the SD radiation energy density predominates and starts to fall below the
magnetic energy density at radii ranging from 20 to 40~kpc, depending on the
particular cluster. Because we are looking at nearby clusters ($z<0.45$) and use
a comparably strong magnetic field, which is in agreement with Faraday rotation
measurements of CCs, the energy density of the CMB never predominates in the
core region ($r<100$~kpc) and remains subdominant out to radii of 200~kpc for
most clusters.

The frequency dependence of the synchrotron emission is encapsulated in $e_{B_\rmn{c}} = B_\rmn{c}^{\C{2}}/(8 \upi)$ where $B_\rmn{c} = 31 \nu \rmn{~GHz^{-1}\umu G}$. The index $\alpha_\nu$ is related to the spectral index of the electron distribution by $\alpha_\nu = (\alpha_\rmn{e} -1)/2$.

The remaining constant $A_\nu$ is given by
\begin{equation}
A_\nu = 4 \upi A_{\rmn{E}_\rmn{synch}} \frac{16^{2-\alpha_\rmn{e}} \sigma_\rmn{pp} m_\rmn{e} c^2}{(\alpha_\rmn{e} - 2) \sigma_\rmn{T} e_{B_\rmn{c}}} \left(\frac{m_\rmn{p}}{m_\rmn{e}} \right)^{\alpha_\rmn{e}-2} \left(\frac{m_\rmn{e} c^2}{\rmn{GeV}}\right)^{\alpha_\rmn{e} - 1}
\end{equation}
with the Thomson cross-section $\sigma_\rmn{T}$ and the effective proton--proton cross-section $\sigma_\rmn{pp}$, which is described as \citep{Pfrommer2004}
\begin{equation}
\sigma_\rmn{pp} = 32 \left(0.96 + \rmn{e}^{4.4-2.4(\alpha_\rmn{e}-1)} \right).\label{eq:sigmapp}
\end{equation}
$A_{\rmn{E}_\rmn{synch}} $ is given by
\begin{equation}
A_{\rmn{E}_\rmn{synch}} = \frac{\sqrt{3 \upi}}{32 \upi} \frac{B_\rmn{c} e^3}{m_\rmn{e} c^2} \frac{\alpha_\rmn{e} + \frac{7}{3}}{\alpha_\rmn{e}+1}\frac{\Gamma\left(\frac{3 \alpha_\rmn{e} -1}{12}\right) \Gamma\left( \frac{3 \alpha_\rmn{e}+7}{12}\right)\Gamma\left(\frac{\alpha_\rmn{e}+5}{4} \right)}{\Gamma\left( \frac{\alpha_\rmn{e}+7}{4}\right)}.
\end{equation}

\subsection{Gamma-ray emission}
\label{sec:appgamma}

The source density for gamma-rays from pion decay as a function of energy is denoted by $s_\gamma (E_\gamma)$. Thus, the number of photons emitted per unit time and area between the energies $E_1$ and $E_2$ is given by
\begin{equation}
\begin{split}
\lambda_{\gamma} &= \int_{E_1}^{E_2} \rmn{d} E_{\gamma} s_{\gamma}(E_{\gamma})\\
&= \frac{4 C_\rmn{p}}{3 \alpha_\rmn{p} \delta_{\gamma}} \frac{m_{\pi^0} c \sigma_\rmn{pp} n_\rmn{N}}{m_\rmn{p}} \left(\frac{m_\rmn{p}}{2 m_{\pi^0}}\right)^{\alpha_\rmn{p}} \left[\mathcal{B}_x \left( \frac{\alpha_\rmn{p}+1}{2 \delta_{\gamma}}, \frac{\alpha_\rmn{p}-1}{1 \delta_{\gamma}}\right) \right]_{x_1}^{x_2}.
\end{split}
\end{equation}
In the last step, we have substituted the source function with the detailed description from \citet{Pfrommer2008}, which assumes that the CR population can be described as in Equation~\eqref{eq:CRpop}. The source function depends primarily on the normalization of the CR population, $C_\rmn{p}(r)$, and the target density, $n_\rmn{N}(r)$. As described in the case of the radio emissivity, we obtain $C_\rmn{p}(r)$ from the CR pressure profile and $n_\rmn{N}(r)$ from fits to observational data. 
We adopt a spectral index for the CR proton population of $\alpha_\rmn{p}=2.4$.
The shape factor $\delta_\gamma$ depends on the spectral index and is given by
\begin{equation}
\delta_\gamma \approx 0.14 \alpha_\rmn{p}^{-1.6} + 0.44.
\end{equation}
The effective proton--proton cross-section $\sigma_\rmn{pp}$ is the same as in Equation~\eqref{eq:sigmapp}. The neutral pion and proton masses are denoted by $m_{\pi^0}$ and $m_\rmn{p}$, respectively.
The last factor contains the incomplete beta function $\mathcal{B}_x(a,b)$ and is evaluated at $x_1$ and $x_2$ with
\begin{equation}
x_i = \left[ 1+\left(\frac{m_{\pi^0} c^2}{2 E_i} \right)^{2 \delta_{\gamma}}\right]^{-1}.
\end{equation}
Luminosities and fluences are obtained via Equations~\eqref{eq:gammal} and \eqref{eq:gammaf}, respectively.

\section{Non-thermal luminosities}
\label{sec:applum}

\begin{figure*}
  \includegraphics{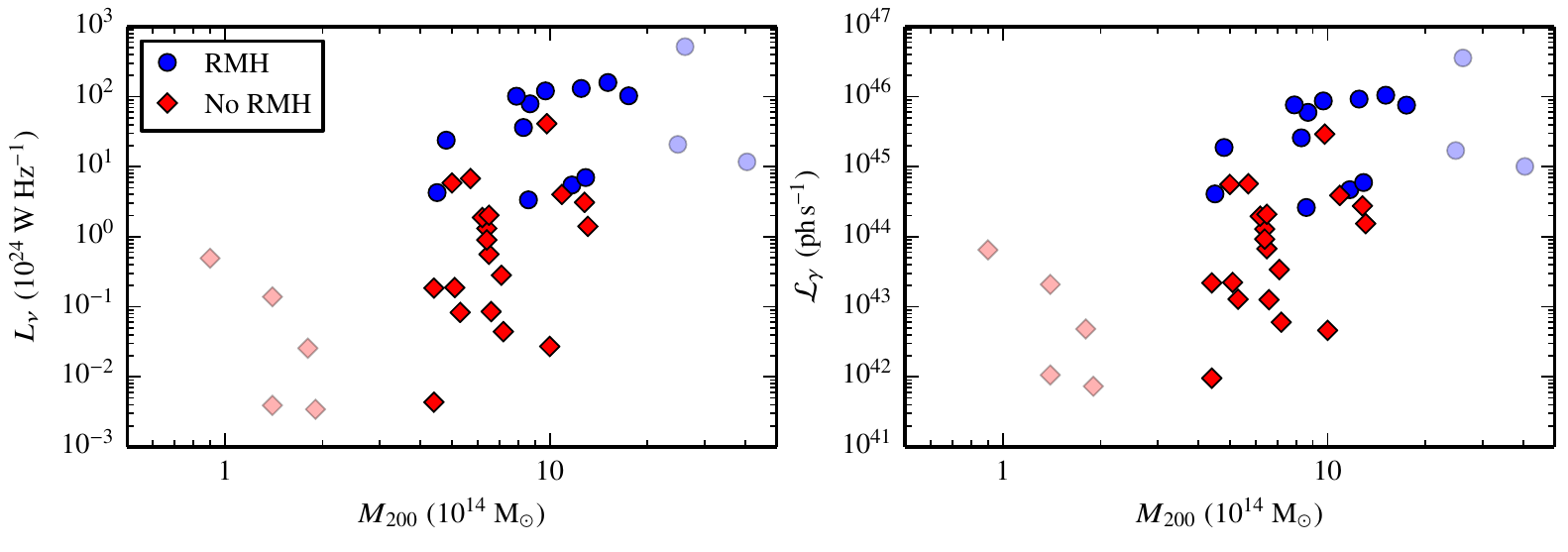}
  \caption{Scaling relations of hadronically induced non-thermal luminosities
    and cluster masses. We show the 1.4~GHz radio luminosity due to secondary
    electrons (left) and the pion-decay gamma-ray luminosity above 1~GeV
    (right). Clusters hosting an RMH (shown with blue circles) populate the
    upper envelope of these relations.  Clearly, both luminosities scale with
    cluster mass. However, there is an enormous scatter in non-thermal
    luminosity at fixed mass due to the large variance in gas density across our
    sample. The latter effect dominates the variance of non-thermal luminosities
    in our core sample (shown with full colours).  }
  \label{fig:LnonthermalvsMass}
\end{figure*}

We show scaling relations of hadronically induced non-thermal luminosities and
cluster masses in Fig.~\ref{fig:LnonthermalvsMass}. We show separately radio
luminosities emitted by secondary CR electrons and gamma-ray emission due to
decaying neutral pions. Assuming that CRs are accelerated at cosmological
structure formation shocks during cosmic history and advectively transported
into clusters, the non-thermal cluster luminosity scales with the virial mass of
clusters as $M_{200}^{\alpha_M}$ with $\alpha_M\approx1.4$ \citep[][excluding
  the signal from the cluster galaxies]{Pfrommer2008_III, Pinzke2011}. We find a
similarly strong scaling with cluster mass. However, this scaling with cluster
mass is accompanied with an enormous scatter in non-thermal luminosity at fixed
mass due to the large variance in gas density across our sample. The latter
effect dominates the variance of non-thermal luminosities in our core sample.

To understand the origin of this scatter, we examine the scaling of the
non-thermal luminosity, $\Lum_\rmn{nt} \propto \int P_\rmn{cr} n f(B) \rmn{d}V =
\int X_\rmn{cr} n^2 kT f(B) \rmn{d}V$, where $f(B)=1$ for the gamma-ray
luminosity and $f(B)$ represents a weak function of magnetic field strength in
the synchrotron-dominated emission regime, i.e., for $e_B\gg e_{\rmn{rad}}$
(equation~\ref{eq:jnu2}). In Fig.~\ref{fig:slopes}, we found a similar spread of
$n$ and $X_\CR$ of a factor of about $30$ in our entire sample. Hence, we expect
$\mathcal{L}_\gamma$ to vary by a factor of about $3\times 10^4$, which is only
marginally reduced to $10^4$ if we restrict ourselves to the core sample,
despite the tight restriction in cluster mass of this subsample.

There is little difference between the relations for the radio and gamma-ray
luminosities, implying that the CR electrons are primarily cooling in the strong
synchrotron regime for which $f(B)$ depends only weakly on magnetic field
strength.  Finally, clusters hosting an RMH populate the upper envelope of these
relations since they signal the CC systems with the highest density (at fixed
radius, see Fig.~\ref{fig:slopes}). The median values of the distribution of RMH
cluster and those without RMHs vary by more than an order of magnitude.

%%%%%%%%%%%%%%%%%%%%%%%%%%%%%%%%%%%%%%%%%%%%%%%%%%

\bsp
\label{lastpage}
\end{document}